\documentclass[12pt]{article}
\usepackage{amsmath}
\usepackage{dsfont}
\usepackage[T1]{fontenc}
\usepackage{fourier}
\usepackage{graphicx}
\usepackage{enumerate}
\usepackage{cite}
\usepackage{rotating}
\usepackage{lscape}
\usepackage{geometry}[margin=1in]
\usepackage{xcolor}
\usepackage{url} 

\definecolor{UBCblue}{rgb}{0.04706, 0.13725, 0.26667}

\begin{document}

\def\spacingset#1{\renewcommand{\baselinestretch}%
{#1}\small\normalsize} \spacingset{1}


\title{\bf Automated calibration of consensus weighted distance-based clustering approaches using sharp}
\author{Barbara Bodinier$^1$, Dragana Vuckovic$^1$, Sabrina Rodrigues$^1$, \\
Sarah Filippi$^2$, Julien Chiquet$^3$ and Marc Chadeau-Hyam$^1$ \vspace{10pt} \\
\small $^1$ Department of Epidemiology and Biostatistics, School of Public Health, Imperial College London, UK. \\
\small $^2$ Department of Mathematics, Imperial College London, London, UK. \\
\small $^3$ Universit\'e Paris-Saclay, AgroParisTech INRAE, UMR MIA, France.}
\maketitle

\setcounter{page}{1}

\begin{abstract}
\textbf{Motivation:} In consensus clustering, a clustering algorithm is used in combination with a subsampling procedure to detect stable clusters. Previous studies on both simulated and real data suggest that consensus clustering outperforms native algorithms. \\

\textbf{Results:} We extend here consensus clustering to allow for attribute weighting in the calculation of pairwise distances using existing regularised approaches. We propose a procedure for the calibration of the number of clusters (and regularisation parameter) by maximising a novel consensus score calculated directly from consensus clustering outputs, making it extremely computationally competitive. Our simulation study shows better clustering performances of (i) models calibrated by maximising our consensus score compared to existing calibration scores, and (ii) weighted compared to unweighted approaches in the presence of features that do not contribute to cluster definition. Application on real gene expression data measured in lung tissue reveals clear clusters corresponding to different lung cancer subtypes. \\

\textbf{Availability and implementation:} The R package sharp (version $\geq$ 1.4.0) is available on CRAN.
\end{abstract}

\noindent%
{\it Keywords:} consensus clustering, calibration, regularisation

\vfill
\hspace{0pt}

\newpage
\spacingset{1.5} 

\section{Introduction}

Clustering aims at partitioning samples (items) into homogeneous groups with similar features (attributes) \cite{FindingGroups}. Distance-based clustering algorithms like hierarchical clustering minimise the pairwise distances within clusters (compactness) and maximise the pairwise distances across clusters (separation). These approaches have been extensively used in medicine, e.g. for stratifying patients based on their molecular profiles \cite{DiseaseSubtyping}. \\

In consensus clustering, a given clustering algorithm is applied on multiple subsamples of items to generate more stable clusters \cite{Monti}. Pairwise co-membership proportions, calculated as the proportions of subsamples for which two items are classified in the same cluster, are stored in the consensus matrix, which is then used as a measure of similarity between the items \cite{Monti}. \\

Consensus clustering has successfully been applied on gene expression data and enabled the identification of stable molecular phenotypes \cite{Adenocarcinoma, SC3}. A simulation study also revealed that consensus clustering is among the best performing clustering approaches when the true number of clusters is known \cite{MPJ}. However the calibration of hyper-parameters hampers its applicability to real-world data sets. \\

Existing methods for the calibration of the number of clusters aim at maximising the stability of the clustering procedure \cite{Monti, ClusteringStability}. It has been proposed to measure clustering stability from the estimated distribution of co-membership proportions by the delta ($\Delta$) score \cite{Monti}, Proportion of Ambiguous Clustering (PAC) score \cite{PAC}, or discrepancy between clustering on the full sample and on perturbed data \cite{DiseaseSubtyping}. More recently, the Relative Cluster Stability Index (RCSI) was introduced as a measure of how much the data can be partitioned and estimated as the difference between the estimated and expected scores (e.g. PAC) under the hypothesis of a single cluster \cite{M3C}. The calculation of the RCSI requires the time-consuming application of consensus clustering on several datasets that are simulated from a reference distribution with a single cluster. \\

In addition, current implementations of consensus clustering with continuous attributes rely on distance metrics defined using vector norms (e.g. Euclidean and Manhattan distances) or correlations (e.g. Pearson's or rank correlation), which assume that all attributes contribute equally to the clustering \cite{rCOSA}. Considering attributes with low proportions of variance explained by the grouping in the distance calculations may dilute the clustering structure, making it more difficult to detect the clusters \cite{COSA, rCOSA}. \\

To overcome this issue, we define here consensus weighted clustering where a distance-based clustering algorithm (e.g. hierarchical clustering) is applied on a weighted distance matrix calculated using existing regularised methods. Specifically, we investigate the use of (i) sparse clustering, which estimates attribute-specific weights that are used in the calculation of pairwise distances between items and can be shrunk to exactly zero for attribute selection \cite{SparseClustering}, and (ii) Clustering Objects on Subsets of Attributes (COSA) algorithm, which estimates attribute and item specific weights \cite{COSA}. \\

We then introduce a novel consensus score measuring the stability of the clustering procedure from (weighted) consensus clustering outputs. We propose to calibrate hyper-parameters by maximising our consensus score. \\

In the Materials and Methods section, we introduce our consensus weighted clustering approach, our calibration procedure and the simulation models implemented. Then, we conduct several simulation studies comparing clustering performances obtained with (i) hierarchical clustering without subsampling, (ii) consensus clustering calibrated as previously proposed \cite{Monti, PAC, M3C}, and (iii) consensus clustering calibrated by maximising our consensus score. We also evaluate both the clustering and ranking performances of consensus weighted clustering. Finally, we apply (consensus weighted) clustering to a publicly available transcriptomics dataset including 3,312 assayed transcripts measured in 17 normal lung tissues and 46 malignant tumours \cite{LungCancerSubtypes}. \\

\section{Materials and methods}

\subsection{Consensus weighted clustering}

\subsubsection{Overview}

\begin{figure}[h!]
\centering
\makebox{\includegraphics[width=0.75\linewidth]{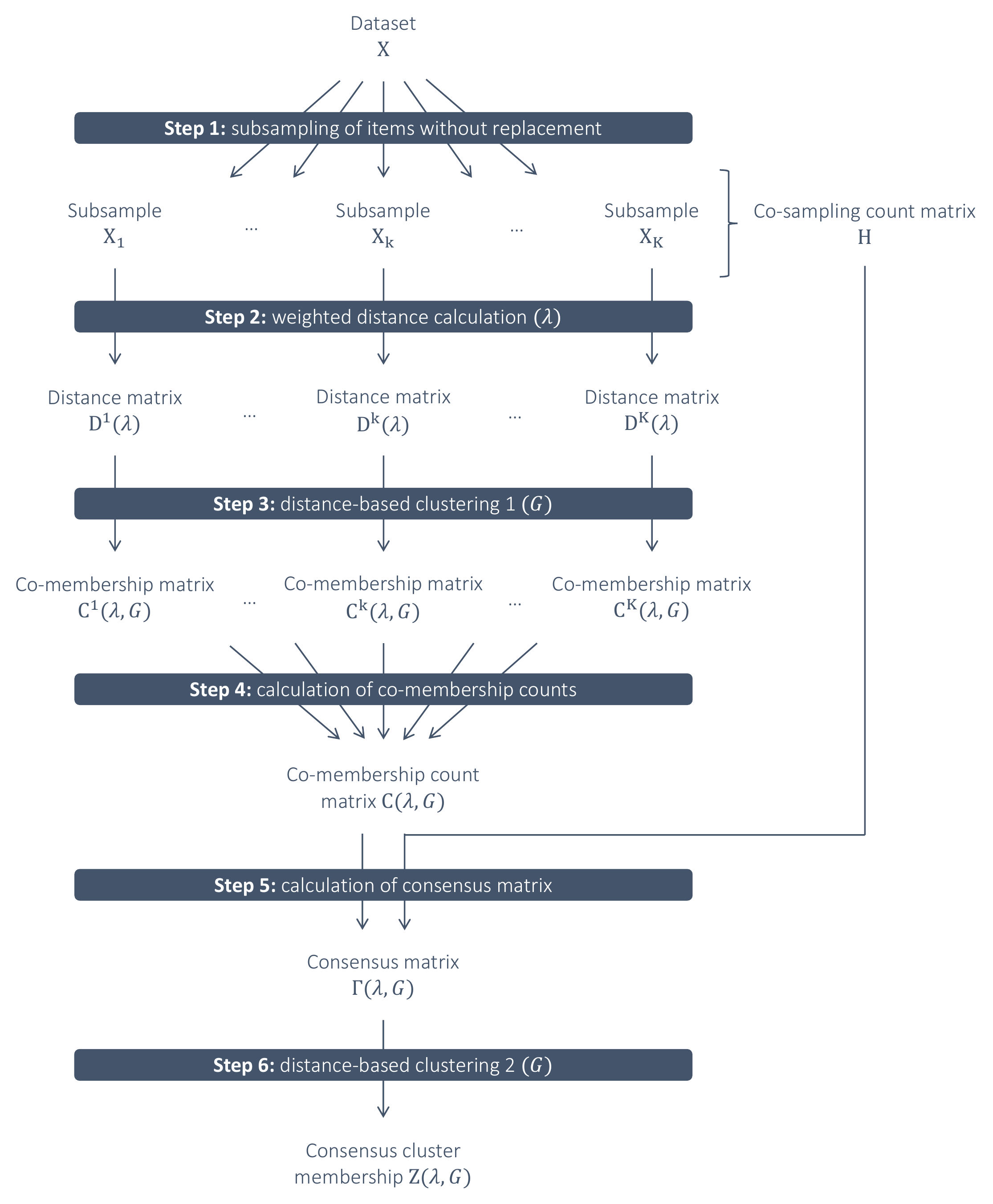}}
\caption{Flowchart describing the six steps of consensus weighted clustering applied on the data matrix $X$ with hyper-parameters $\lambda$ and $G$.}
\label{fig:steps}
\end{figure}

The proposed consensus weighted clustering approach can be decomposed into 6 steps, as described in Figure \ref{fig:steps}. Consensus weighted clustering is governed by two hyper-parameters that need to be calibrated: the regularisation parameter $\lambda$ for the calculation of weighted distances and the number of clusters $G$. 

First, $K$ random subsamples of a proportion $\tau \in [0,1]$ of the $n$ items are drawn without replacement. We use $K=100$ and $\tau=0.5$ throughout this paper, as in previous studies \cite{M3C, arxiv}. The number of subsamples where each pair of items is drawn is stored in the co-sampling count matrix $H$. Second, a (weighted) distance matrix is calculated for each of the $K$ subsamples with parameter $\lambda$. Third, a distance-based clustering algorithm is applied on each of the $K$ distance matrices to detect $G$ clusters. Based on the resulting cluster memberships in each of the $K$ subsamples, co-membership matrices indicating which pairs of items belong to the same cluster are calculated. Fourth, the matrix of co-membership counts $C(\lambda, G)$ is calculated as the sum over the $K$ co-membership matrices. Fifth, the consensus matrix $\Gamma(\lambda, G)$ is calculated and contains, for each pair of items, the proportion of co-membership out of the number of subsamples where both items were drawn. \\

Consensus (weighted) clustering aims at the identification of stable clusters by maximising the within-cluster and minimising the between-cluster co-membership proportions obtained over multiple subsamples. In Step 6, a distance-based clustering algorithm (potentially different from the one in Step 3) is used to generate the $G$ stable clusters in $Z(\lambda, G)$ using the consensus matrix as a measure of similarity. \\

These steps are described in detail in the following sections. 

\subsubsection{Weighted distance calculation}

We extend the consensus clustering framework by incorporating an algorithm for weighted distance matrix calculation in Step 2 (Figure \ref{fig:steps}). We investigate the use of two algorithms: sparse clustering \cite{SparseClustering} and Clustering Objects on Subsets of Attributes \cite{COSA}. 

\paragraph{Sparse clustering} 

The sparse clustering approach introduced in \cite{SparseClustering} aims at identifying clusters that are supported by a subset of discriminatory attributes. This is achieved by introducing $p$ attribute-specific weights $w_m, m \in \{1, \dots, p\}$ in the distance calculations \cite{SparseClustering}. For given attribute weights $w$, the sparse clustering distance $d_{ij}^S (w)$ between items $i$ and $j$ can be expressed as:

\begin{equation}
d_{ij}^S (w) = \sum_{m = 1}^p w_m d_{ijm}
\end{equation}
where $d_{ijm}$ is the pairwise distance along attribute $m$. \\

Attribute weights are estimated by solving a regularised version of the clustering objective function. In the present paper, we use the weighted distance matrix corresponding to sparse hierarchical clustering \cite{SparseClustering}:
\begin{align}
\max_{w, U} & \sum_{m = 1}^p w_m \left[ \sum_{i=1}^n \sum_{j=1}^n \left( d_{ijm} U_{ij} \right) \right], \text{ such that}  \nonumber \\
& \sum_{ij} U_{ij}^2 \leq 1, ||w||_2^2 \leq 1, ||w||_1 \leq \lambda, w_m \geq 0, \forall m \in \{1, \dots, p\} \nonumber
\end{align}
where $n$ is the number of items to cluster and $U$ is the overall distance matrix. \\

The constraint on the $\ell_1$-norm of $w$ by a regularisation parameter $\lambda > 1$ induces sparsity, i.e. results in some weights $w_m$ being shrunk to exactly zero. The use of the $\ell_2$-norm of $w$ ensures that the clustering is not driven by a single feature. The calibration of the regularisation parameter $\lambda$ conditionally on the number of clusters can be done using an adapted gap statistic measuring the difference between the observed and expected objectives assuming a single cluster. 

\paragraph{Clustering Objects on Subsets of Attributes}

It has been shown that the estimation of a single weight per attribute may result in clusters that are equally spaced along the set of selected features \cite{COSA, rCOSA}. Clusters that are supported by cluster-specific attributes may therefore be missed. An alternative approach introduces attribute and item specific weights in Clustering Objects on Subsets of Attributes (COSA) \cite{COSA}. In COSA, the weight matrix $W$ of size $( n \times p)$ is estimated by minimising the sum of weighted distances between each item and its nearest neighbours under a constraint on the weights \cite{COSA}. Entries of the COSA weighted distance matrix $d^C (W)$ are given by
\[
d_{ij}^C (W) = \sum_{m=1}^p \max(W_{im}, W_{jm}) d_{ijm}
\]

The $(n \times p)$ weights in $W$ are estimated from
\[
\min_{W} \sum_{i=1}^n \left[ \frac{1}{\sqrt{n}} \sum_{j \in KNN(i)} d_{ij} (W_{i.}) + \lambda \sum_{m=1}^p W_{im} \log (W_{im}) \right]
\]
where $KNN(i)$ are the $\sqrt{n}$ nearest neighbours of item $i$, $W_{i.}$ is the $i^{th}$ row of matrix $W$ and $d_{ij} (W_{i.}) = \sum_{m=1}^p W_{im} d_{ijm}$.

This optimisation problem is solved approximately using an iterative algorithm. The amount of regularisation is controlled by the parameter $\lambda$ but does not result in attribute selection.

\subsubsection{Consensus clustering framework}

In Step 3 (Figure \ref{fig:steps}), a distance-based clustering algorithm is applied on each of the $K$ (weighted) distance matrices. The co-membership status $C_{ij}^k (\lambda, G)$ of items $i$ and $j$ in subsample $k$ is defined as:

\[
C_{ij}^k (\lambda, G) =
\begin{cases}
1 \text{ \small if $i \neq j$ are both in subsample $k$ and are in the same cluster obtained with parameters $\lambda$ and $G$,} \\
1 \text{ \small if $i = j$,} \\
0 \text{ \small otherwise.}
\end{cases}
\]

The matrix $C$ of co-membership counts over the $K$ iterations is then calculated in Step 4 (Figure \ref{fig:steps}) as:

\[
C_{ij} (\lambda, G) = \sum_{k = 1}^K C_{ij}^k (\lambda, G)
\]

To account for the possibility that items $i$ and $j$ may not both be included in a given subsample, we define the matrix $H$ of co-sampling counts, where $H_{ij}$ 
is the number of subsamples that include both items $i$ and $j$. By construction, the element $H_{ii}$ corresponds to the number of subsamples that include item $i$. We consider throughout this paper that the same subsamples are used to calculate the co-membership counts for all pairs of parameters $(\lambda, G)$. That is, the matrix $H$ does not depend on $\lambda$ or $G$. \\

In Step 5 (Figure \ref{fig:steps}), entries of the consensus matrix $\Gamma (\lambda, G)$ are defined as co-membership proportions calculated over subsamples where both $i$ and $j$ were included:

\[
\Gamma_{ij} (\lambda, G) = \frac{C_{ij} (\lambda, G)}{H_{ij}}
\]

As previously proposed \cite{Monti, PAC, M3C}, the $G$ stable clusters in $Z(\lambda, G)$ are obtained by applying a distance-based clustering algorithm using co-membership proportions as a measure of pairwise similarity (Step 6 in Figure \ref{fig:steps}). The use of co-membership proportions instead of the Euclidean distance in the construction of stable clusters may result in the re-assignment of some items (Supplementary Figure 1). \\

Note that consensus \textit{unweighted} clustering is recovered by setting $\lambda = 0$.

\subsection{Calibration of hyper-parameters}

\subsubsection{Existing scores}

To select the number of clusters $G$, it has been proposed in \cite{Monti, ClusteringStability} to compare consensus matrices $\Gamma (G)$ obtained with different values of $G$ and select the one yielding the most stable clustering. Theoretically, the most stable clustering would invariably result in the same partition at all subsampling iterations, yielding a binary consensus matrix. Based on this, the proportion of item pairs with co-membership proportions above a certain threshold $x$ was proposed as a measure of stability \cite{Monti, CCPlus}:

\begin{equation}
CDF_{G} (x) = \frac{1}{n (n-1) /2} \sum_{i<j} \mathds{1}_{\Gamma_{ij} (G) \leq x}.
\end{equation}

Using that metric, Monti et al. (2003) propose to determine the number of clusters $G$ as the value generating the largest standardised score $\Delta_{G}$:

\begin{equation}
\Delta_{G} = 
\begin{cases}
a_{G} & \text{if } G = 2.\\
( a_{G} - a_{G -1} ) / a_{G-1} & \text{if } G > 2, 
\end{cases} \nonumber
\end{equation}
where $a_G$ is the integral of $CDF_G (x)$ over $[ 0, 1 ]$ and increases as co-membership proportions get larger. \\

Alternatively, the Proportion of Ambiguous Clustering (PAC) score measures the number of intermediate co-membership proportions (i.e. that are between the lowerbound $x_1$ and upperbound $x_2$ of the co-membership proportions) and is defined as\cite{PAC}: 

\[
\text{PAC}_{G} (x_1, x_2) = \text{CDF}_{G} (x_2) - \text{CDF}_{G} (x_1).
\]

The number of clusters $G$ can then be defined as the one yielding the smallest PAC score (i.e. with the smallest number of intermediate co-membership proportions). The calculation of the PAC score requires the arbitrary choice of two parameters $x_1$ and $x_2$. Default values of $x_1 = 0.1$ and $x_2 = 0.9$ were recommended \cite{PAC}. \\

A score measuring the discrepancy between clustering on the full sample and on perturbed datasets has been proposed to choose the number of clusters in the Perturbation clustering for data INtegration and disease Subtyping (PINS) algorithm \cite{DiseaseSubtyping, PINSPlus}. The discrepancy score measures clustering stability as the Area Under the Curve (AUC) of the Cumulative Distribution Function (CDF) of the difference between the co-membership matrix obtained by applying the clustering algorithm on the full sample and the consensus matrix. \\

More recently, a Monte Carlo reference-based technique was proposed and implemented in the R package \texttt{M3C} \cite{M3C}. The approach is based on the simulation of multiple (package default is $25$) datasets from a reference distribution with a single cluster while keeping a similar attribute correlation structure \cite{GAP}. The number of clusters is then calibrated by maximising the Relative Cluster Stability Index (RCSI) of the PAC score or entropy \cite{M3C, entropy} measuring the difference between observed and simulated scores. 

\subsubsection{New consensus score}

In this section, we introduce a novel consensus score $S_c$ measuring clustering stability and used for the joint calibration of hyper-parameter(s). The consensus score $S_c$ is calculated using the matrix $C(\lambda, G)$ of co-membership counts, the matrix $H$ of co-sampling counts and the consensus clusters $Z(\lambda, G)$, which are all outputs of consensus clustering (Figure \ref{fig:steps}). The co-membership count matrix $C(\hat{\lambda}, \hat{G})$ and consensus clusters $Z(\hat{\lambda}, \hat{G})$ obtained with the calibrated number of clusters $\hat{G}$ and penalty parameter $\hat{\lambda}$ (if weighted) maximise the score $S_c$. \\

\begin{figure}[h!]
\centering
\makebox{\includegraphics[width=\linewidth]{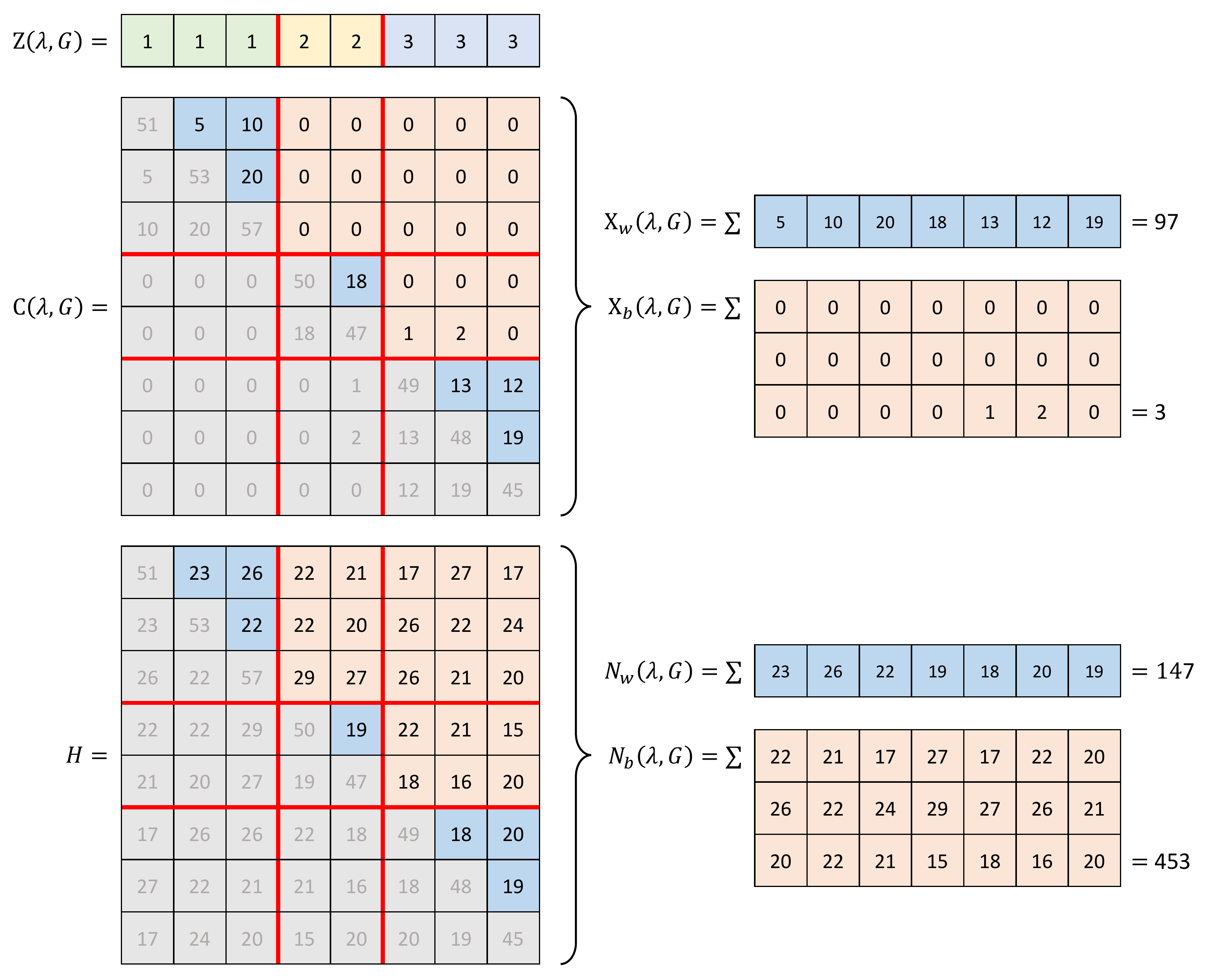}}
\caption{Illustration of the calculation of the quantities $X_w(\lambda, G)$, $X_b(\lambda, G)$, $N_w(\lambda, G)$, and $N_b(\lambda, G)$ given the consensus clustering outputs $Z (\lambda, G)$, $C (\lambda, G)$, and $H$. The blue entries indicate the \textit{within} elements and the orange entries are \textit{between} elements.}
\label{fig:score}
\end{figure}

To calculate the consensus score for a given pair of hyper-parameters $(\lambda, G)$, the $N = n \times (n-1) /2$ item pairs are first classified as (i) \textit{within} elements if the two items belong to the same consensus cluster as defined in $Z (\lambda, G)$, or (ii) \textit{between} elements if the two items belong to different consensus clusters. This is illustrated in Figure \ref{fig:score}, where the blue and orange entries of the co-membership count matrix $C (\lambda, G)$ and co-sampling count matrix $H$ correspond to the within and between elements, respectively. \\

\pagebreak

We introduce the integers $X_w(\lambda, G)$, $X_b(\lambda, G)$, $N_w(\lambda, G)$ and $N_b(\lambda, G)$ computed as follows (Figure \ref{fig:score}):
    \begin{align}
        X_w(\lambda, G) & = \sum_{i < j} C_{ij} (\lambda, G) \mathds{1}_{Z_i (\lambda, G) = Z_j (\lambda, G)} \\
        X_b(\lambda, G) & = \sum_{i < j} C_{ij} (\lambda, G) \mathds{1}_{Z_i (\lambda, G) \neq Z_j (\lambda, G)} \\
        N_w(\lambda, G) & = \sum_{i < j} H_{ij} \mathds{1}_{Z_i (\lambda, G) = Z_j (\lambda, G)} \\
        N_b(\lambda, G) & = \sum_{i < j} H_{ij} \mathds{1}_{Z_i (\lambda, G) \neq Z_j (\lambda, G)}
    \end{align}

The quantity $X_w(\lambda, G)$ is the total number of co-members obtained over the $K$ subsampling iterations that are among the \textit{within} pairs of items (in blue in Figure \ref{fig:score}). The number $N_w (\lambda, G)$ is the total number of \textit{within} pairs of items that are sampled together over the $K$ subsampling iterations. Similarly, $X_b(\lambda, G)$ and $N_b (\lambda, G)$ are the total numbers of co-members and of co-sampled pairs, respectively, among the \textit{between} pairs (in orange in Figure \ref{fig:score}). \\

Note that the co-membership counts $C_{ij}(\lambda, G)$ follow independent binomial distributions, conditionally on the co-sampling counts $H$ and consensus clusters $Z(\lambda, G)$:
\[
C_{ij} (\lambda, G) | H, Z(\lambda, G) \sim \mathcal{B} \left( H_{ij}, p_{ij}(\lambda, G) \right)
\]
Our consensus score evaluates if the probabilities $p_{ij} (\lambda, G)$ are larger for pairs of items in the same consensus cluster ($Z_i = Z_j$) than for pair of items in different consensus clusters ($Z_i \neq Z_j$). To devise this score, we assume for simplicity that 
$$p_{ij}(\lambda, G) = 
\begin{cases}
p_w(\lambda, G) \text{ if $Z_i = Z_j$} \\
p_b(\lambda, G) \text{ otherwise.}
\end{cases}.$$
As a consequence, the quantities $X_w (\lambda, G)$ and $X_b (\lambda, G)$ also follow binomial distributions, conditionally on the co-sampling counts in $H$:
\[
X_w (\lambda, G) | H, Z (\lambda, G) \sim \mathcal{B} \left( N_w (\lambda, G), p_w (\lambda, G) \right) 
\text{ and } 
X_b (\lambda, G) | H, Z (\lambda, G) \sim \mathcal{B} \left( N_b (\lambda, G), p_b (\lambda, G) \right)
\]

We consider that a stable clustering is characterised by a probability $p_w (\lambda, G)$ that is larger than $p_b (\lambda, G)$. To measure clustering stability, we then compare the probabilities $p_w (\lambda, G)$ and $p_b (\lambda, G)$ using a two-sample z test where the null hypothesis is $p_w (\lambda, G) \leq p_b (\lambda, G)$. The consensus score $S_c$ is defined as the z statistic, calculated as:

\[
S_c (\lambda, G) = \frac{\hat{p}_w (\lambda, G) - \hat{p}_b (\lambda, G)}{\sqrt{\hat{p}_0 (\lambda, G) \left( 1 - \hat{p}_0 (\lambda, G) \right) \left( \frac{1}{N_w (\lambda, G)} + \frac{1}{N_b (\lambda, G)} \right)}}
\]

where $\hat{p}_w (\lambda, G) = \frac{X_w (\lambda, G)}{N_w (\lambda, G)}$, $\hat{p}_b (\lambda, G) = \frac{X_b (\lambda, G)}{N_b (\lambda, G)}$, and $\hat{p}_0 (\lambda, G) = \frac{X_w (\lambda, G) + X_b (\lambda, G)}{N_w (\lambda, G) + N_b (\lambda, G)}$. \\

For large enough $X_w (\lambda, G)$ and $X_b (\lambda, G)$ (typically $N_w (\lambda, G) p_w (\lambda, G) > 10$ and $N_b (\lambda, G) p_b (\lambda, G) >10$), the z statistic approximately follows a standard Normal distribution \cite{NormalApprox, StatsAndProbaBook}. As such, the consensus score $S_c (\lambda, G)$ is comparable across different numbers of clusters $G$ and penalty parameters $\lambda$. The consensus score $S_c (\lambda, G)$ increases with clustering stability. \\

To illustrate this score, we now consider the extreme situation where the clustering is the most stable. The most stable clustering would result in a binary consensus matrix, or equivalently in a matrix $C(\lambda, G)$ where (i) \textit{within} elements are the same as in $H$, and (ii) \textit{between} elements are all zero. In this extreme situation, the z statistic is equal to $z (\lambda, G) = \sqrt{N_w (\lambda, G) + N_b (\lambda, G)} = \sqrt{N}$. This is the maximum value that the z statistic can take, which results in the maximum value of the stability score (see Appendix for proof). In addition, the z statistic does not depend on $\lambda$ or $G$ in this situation. This is a desirable property as it implies that the most stable clustering (associated with a binary consensus matrix) would result in the same consensus score regardless of the numbers and sizes of clusters.

\subsubsection{Grid search}

The number of clusters $G$ (and regularisation parameter $\lambda$ for weighted clustering) are calibrated by maximising the consensus score $S_c$ using a grid search algorithm where the consensus matrix and metric measuring the stability are computed for different values of $G$ (and $\lambda$). The calibrated (set of) parameter(s) is the one that maximises our consensus score.

\subsection{Simulation models}

\subsubsection{Gaussian mixture model}

We simulate data $X$ including $n$ items and $p$ attributes from a Gaussian mixture, $\mathcal{M}$, where, $\forall i \in \{1, \dots, n\}$ \cite{GMM}:
\begin{equation}
\begin{split}
Z_i \text{ i.i.d. } & \sim \mathcal{M} \left( 1, \kappa \right) \\
X_i | Z_i \text{ independent } & \sim \mathcal{N}_p \left( \mu_{Z_i}, \Sigma \right)
\end{split}
\label{eq:simul}
\end{equation}

where $\kappa$ is a vector of length $G$ of probabilities that an item belongs to a given cluster, $\mu_{Z_i}$ is the mean vector of length $p$ for item $i$ belonging to cluster $Z_i$ and $\Sigma$ is the covariance matrix of size $(p \times p)$. \\

Items belonging to different clusters are generated using the same covariance matrix $\Sigma$ but different mean vectors $\mu_{Z_i}$. The number and size of the simulated clusters is controlled via the number of entries in vector $\kappa$. We directly use the vector of true cluster membership $Z$ as a simulation parameter in our simulations. \\

\subsubsection{Simulation of cluster means $\mu_g$}

To control the level of cluster separation and compactness by attribute, we first sample intermediate cluster- and attribute-specific means $\eta_{gj}$ for each cluster $g \in \{1, \dots, G\}$ and each attribute $j \in \{1, \dots, p\}$ from a Gaussian distribution with mean zero and unit variance. The $(G \times p)$ intermediate cluster- and attribute-specific means are then stored in the matrix $\tilde{M}$ of size $(n \times p)$ such that $\tilde{M}_{ij} = \eta_{Z_i, j}$. \\

The level of separation between the clusters along attribute $j$ is controlled by the expected proportion of explained variance $E_j \in [0, 1]$. To ensure that the desired proportion of explained variance is reached, we generate the mean $M_{ij}$ for item $i$ along attribute $j$ using:

\[ 
M_{ij} = \frac{\sqrt{E_j} \times ( \tilde{M}_{ij} - \frac{1}{n} \sum_{k = 1}^p \tilde{M}_{ij} )}{\sqrt{\frac{1}{n - 1} \sum_{i = 1}^n \left( \tilde{M}_{ij} - \frac{1}{n} \sum_{k = 1}^n \tilde{M}_{kj} \right)^2}}, 
\]

where $E_j$ is the desired proportion of variance along attribute $j$ explained by the simulated clustering. \\

The $i$-th row of the simulated matrix $M$ is the mean vector $\mu_{Z_i}$ and can directly be used in the simulation model presented in Equation \ref{eq:simul}. \\

The random sampling of cluster means introduces some variability in the level of separation between pairs of clusters along a given attribute (Supplementary Figure 2). By aggregating information over multiple attributes, the distances calculated for the simulated data are overall lower for pairs of items belonging to the same cluster than for pairs of items belonging to different clusters. The level of separation and compactness of the clusters is controlled by the number of attributes $p$ and the proportions of explained variance by attribute $E_j, j \in \{1, \dots, p\}$ (Supplementary Figure 3). \\

By chance, we may obtain some clusters that are very well separated from all others, and sets of clusters that are not as well separated (Supplementary Figure 4) \cite{silhouette}. For example, in Supplementary Figure 4, cluster 4 (in red) is very compact and well separated from other clusters, as indicated by the large silhouette widths for all cluster members. On the other hand, clusters 3 and 5 (yellow and green) include items with silhouette widths below zero indicating poor cluster separation. The proposed simulation procedure controls the overall cluster separation through the proportion of variance explained by the grouping structure for each attribute. \\


\subsubsection{Simulation of covariance matrix $\Sigma$}

We first simulate a correlation matrix $\tilde{\Sigma}$. We consider two simulation scenarios with (i) independent attributes, or (ii) groups of correlated attributes. In the first scenario, the identity matrix is used as correlation matrix. In the second scenario, the correlation matrix $\tilde{\Sigma}$ is simulated as previously proposed in the context of graphical modelling \cite{arxiv, Rfake}. Briefly, we (i) simulate the adjacency matrix of a graph with connected components of random subgraphs, (ii) simulate a corresponding precision matrix, (iii) invert it to obtain a covariance matrix, and (iv) compute the correlation matrix $\tilde{\Sigma}$ from the covariance matrix. \\

To ensure that the proportion of variance explained by the grouping for variable $X_j$ is equal to $E_j$, the covariance matrix $\Sigma$ used in Equation \ref{eq:simul} is defined as:

\[
\Sigma_{ij} = \sqrt{(1 - E_i) \times (1 - E_j)} \times \tilde{\Sigma}_{ij}
\]

This simulation model has been implemented in the R package \texttt{fake} (version $\geq$ 1.4.0), available on CRAN \cite{Rfake}.

\subsection{Performance metrics}

Clustering performance is measured by the Adjusted Rand Index (ARI), calculated by comparing the true and estimated co-memberships \cite{Rand, ARI}: 

\[
\text{ARI} = \frac{2 \times (TP \times TN - FP \times FN)}{(TP+FP) \times (TN+FP)+(TP+FN) \times(TN+FN)}
\]

where $TP$ is the number of True Positives (i.e. true co-members that are in the same reconstructed clusters), $TN$ is the number of True Negatives (i.e. pairs of items that are correctly put in different clusters), and $FN$ is the number of False Negatives (i.e. true co-members that are in different reconstructed clusters). \\

Feature selection performance is measured by the $F_1$-score \cite{InformationRetrieval}:

\[
F_1=\frac{2 \times P \times R}{P + R}
\]

where $P$ is the precision and $R$ is the recall calculated by comparing the sets of attributes contributing to the clustering in the simulation (i.e. such that $E_j \neq 0$) and attributes with the highest median weights estimated for weighted distances.

\section{Results}

\subsection{Simulation study}

\subsubsection{Outline}

In this section, we apply (consensus) clustering to simulated datasets with different numbers of items, attributes, clusters and levels of separation. Unless specified otherwise, inferences are based on hierarchical clustering with complete linkage applied to the (weighted) Euclidean distance. We use $K=100$ subsampling iterations throughout this paper. First, we compare the clustering performances of hierarchical clustering and consensus unweighted clustering calibrated using different strategies. Then, we evaluate both the clustering and weighting performances of consensus weighted clustering. \\

For unweighted clustering, comparisons are conducted using $1,000$ simulated datasets with $p = 10$ attributes and $n = 150$ items allocated to $G^* = 5$ clusters of sizes $N_1 = 20$, $N_2 = 50$, $N_3 = 30$, $N_4 = 10$, $N_5 = 40$. The $p$ attributes all have the same proportion $E$ of explained variance by the clustering. Different levels of cluster separation are investigated with $E$ ranging from $0.4$ to $0.6$. Sensitivity analyses include different numbers and sizes of clusters. \\

For weighted clustering, a total of $p = 100$ independent attributes are simulated, of which $q^* = 20$ have nonzero proportions of explained variance ($E = 0.6$). Sensitivity analyses use different numbers of contributing attributes $q^*$, different proportions of explained attribute variances by the clustering and/or groups of correlated attributes.

\subsubsection{Comparison of calibration scores}

We represent clustering performance as measured by the Adjusted Rand Index (ARI) obtained with the existing and novel calibration scores for different numbers of clusters (Supplementary Figure 5). The simulated number of clusters ($G^* = 5$) is recovered using all scores except for the $\Delta$ score ($\hat{G}_{\Delta} = 2$). We generally observe increasing ARI with the RCSI and consensus scores, suggesting that these metrics are relevant to choose the model with the best clustering performance (Supplementary Figure 5E-G). \\

Evaluating the calibration performances of these approaches for different values of separation and compactness, we found that the PAC score is only able to detect the correct number of simulated clusters for well separated clusters (Supplementary Figure 6A-B). The correct number of clusters is always missed by the $\Delta$ score in these examples. The RCSI and consensus scores are able to detect the correct number of clusters even under limited levels of separation (Supplementary Figure 6C). When the clustering structure is not detected, the RCSI scores suggest small numbers of clusters ($\hat{G}_{RCSI} = 2$) while the consensus score chooses a large number ($\hat{G}_{S_c} = 18$) (Supplementary Figure 6E).

\subsubsection{Clustering performance}

\begin{figure}[h!]
\centering
\makebox{\includegraphics[width=\linewidth]{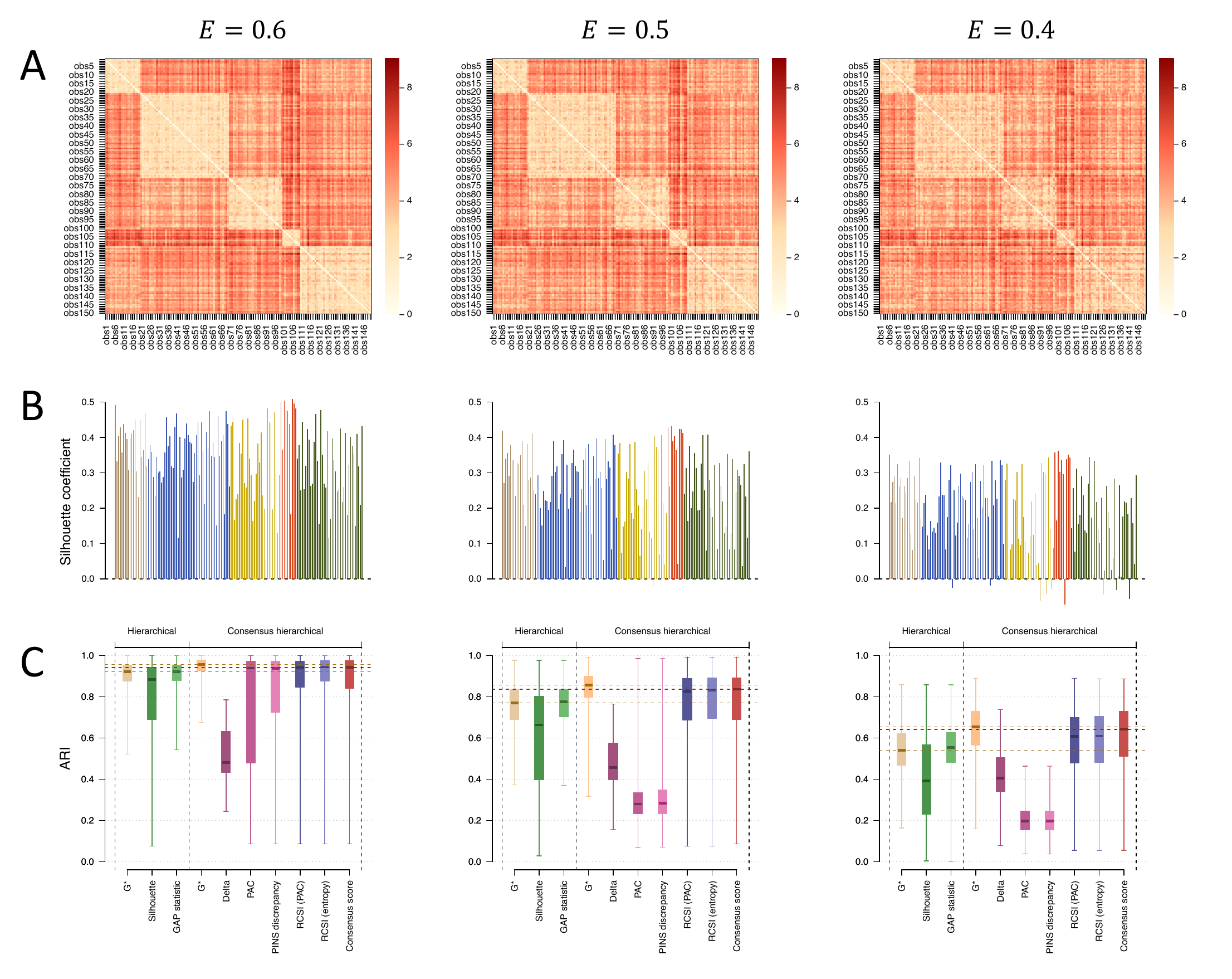}}
\caption{Comparison of clustering performances of (consensus) hierarchical clustering with different calibration strategies from $N=1,000$ simulated datasets corresponding to different levels of cluster separation. We simulate $N=1,000$ datasets with $n=150$ items split into $G^* = 5$ clusters such that $N_1 = 20$, $N_2 = 50$, $N_3 = 30$, $N_4 = 10$, $N_5 = 40$ across $p=10$ features, each with a proportion of explained variance of $E=0.6$ (left), $E=0.5$ (middle) or $E=0.4$ (right). For each scenario, we show a heatmap of Euclidean distances (A) and barplot of silhouette widths for each of the $n = 150$ items coloured by simulated cluster membership (B) for one simulated dataset. Median, quartiles, minimum and maximum Adjusted Rand Index (ARI) for hierarchical clustering with the simulated number of clusters ($G^*$), or calibrated by maximising the silhouette and GAP score, and for consensus hierarchical clustering with $G^*$ or calibrated using the $\Delta$, PAC, PINS discrepancy, RCSI and consensus scores are reported (C).}
\label{fig:performances}
\end{figure}

We evaluate the ability of hierarchical and consensus clustering with different calibration strategies to recover the grouping structure in simulated data with different levels of cluster separation (Figure \ref{fig:performances}). \\

As a reference, we report the performances of both clustering models with the number of clusters used for the simulation $G^* = 5$ and compare them with results based on the calibrated numbers of clusters. With the true number of clusters $G^*$, consensus clustering outperforms hierarchical clustering in the three settings, with a larger increase in ARI for weaker levels of cluster separation (Figure \ref{fig:performances}). This can be explained by the re-assignment of items (mostly re-attributed to their correct cluster) when using the consensus matrix as a measure of similarity (Supplementary Figure 1). \\

For hierarchical clustering, calibration maximising the GAP statistic \cite{GAP} generates better performances than with the silhouette coefficient \cite{silhouette, cluster} in all three scenarios (Figure \ref{fig:performances}). \\

For consensus clustering, models calibrated by maximising the $\Delta$ score perform poorly in the three settings (Figure \ref{fig:performances}). We observe good performances of calibration by the PAC score for well separated clusters only ($E = 0.6$), but the inter-quartile range of the ARI is larger than for other approaches (Supplementary Table 1). Models calibrated using the RCSI and consensus scores generate the best clustering performances in these scenarios and are able to recover the true number of clusters $G^*$ for most simulated datasets (Supplementary Table 1). For weaker cluster separation ($E=0.4$), calibration by maximising the consensus score outperforms all other approaches, including the Monte Carlo based procedures (increase in median ARI of $0.03$, Supplementary Table 1). This suggests that our score allows for the detection of more subtle clustering structures. Increasing the number of Monte Carlo sampling iterations from 25 to 100 does not affect the clustering performances of models calibrated using the RCSI scores (Supplementary Table 2). \\

The estimation of consensus matrices for 2 to 20 clusters on these simulated datasets took less than 5 seconds using 1 CPU and 1 GB memory (Supplementary Table 1). The time needed to compute the silhouette, $\Delta$, PAC or consensus scores is less than a second. The GAP statistics were calculated in 1-2 seconds. The median time to calculate the RCSI scores was above 2 minutes for $N = 25$ Monte Carlo sampling iterations (Supplementary Table 1). \\

Overall these results suggest that the proposed score performs at least as well as established approaches and generates performances that are very close to those of consensus clustering using the true number of clusters $G^*$ in all scenarios (difference in median ARI lower than $0.02$), for no increase in computation time once consensus matrices are estimated (Supplementary Table 1). Conclusions are similar when applying these approaches using partitioning around medoids (Supplementary Figure 7) and for simulated data with different numbers of items (Supplementary Table 3) or clusters (Supplementary Figure 8).

\subsubsection{Performance of consensus weighted clustering}

\begin{figure}[h!]
\centering
\makebox{\includegraphics[width=0.8\linewidth]{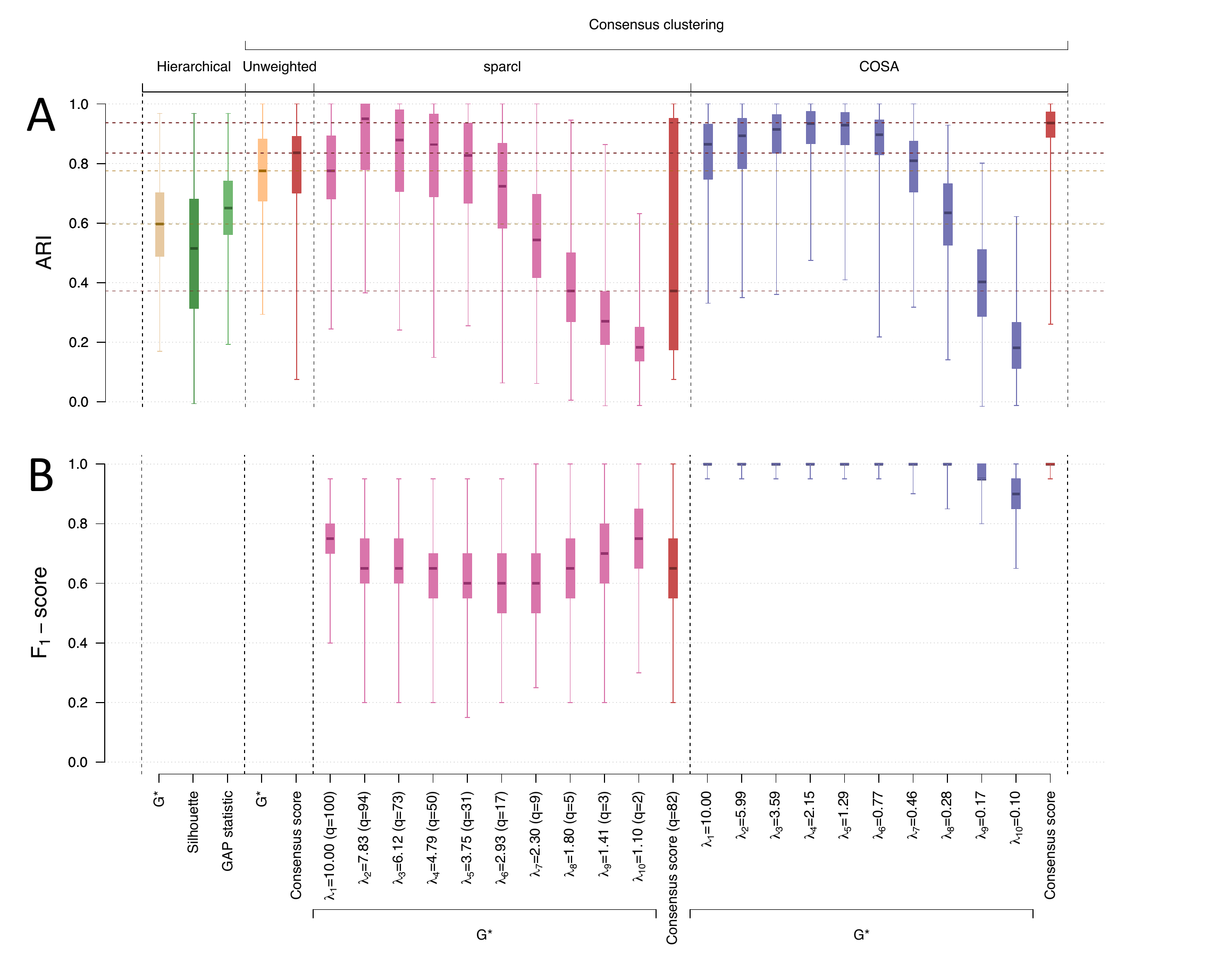}}
\caption{Comparison of clustering performances by the Adjusted Rand Index (ARI) (A) and of the attribute weighting (when applicable) by the $F_1$-score (B) for hierarchical and consensus clustering using unweighted, sparcl or COSA Euclidean distances. Performances are evaluated for $N=1,000$ simulated datasets with $n=150$ items split into $G^*=5$ clusters including $N_1 = 20$, $N_2 = 50$, $N_3 = 30$, $N_4 = 10$, $N_5 = 40$ items, respectively. The clustering structure is supported by $q^* = 20$ of the $p = 100$ attributes with nonzero proportion of explained variance ($E=0.6$). Hierarchical clustering is conducted using a number of clusters that is simulated ($G^*$) or calibrated using the silhouette coefficient or the GAP statistic. For unweighted consensus clustering, the number of clusters is the simulated ($G^*$) or calibrated number maximising the consensus score. For weighted clustering using sparcl or COSA distances, we (i) fix the number of clusters to $G^*$ and consider ten different values of $\lambda$, or (ii) jointly calibrate the number of clusters and the penalty parameter using our consensus score (A). For consensus weighted clustering, the $F_1$-score measures weighting performance by comparing the sets of (i) the 20 features with highest weights (or selection proportions for sparcl), and (ii) the 20 features supporting the clustering in the simulation (B). }
\label{fig:weighting}
\end{figure}

We simulate data with $G^*=5$ clusters that can be observed along $q^* = 20$ of the $p=100$ attributes (i.e. with nonzero proportion of explained variance by the grouping). We use implementations in the R packages sparcl for hierarchical sparse clustering and rCOSA for Clustering Objects on Subsets of Attributes. \\

Using the true number of clusters $G^*=5$, we observe an increase in clustering performance when introducing weighting in the algorithm with a median ARI of 0.78 for (unweighted) consensus clustering compared to 0.95 and 0.94 for the best sparcl and COSA models, respectively (Figure \ref{fig:weighting}A, Supplementary Table 4). \\

Joint calibration of the number of clusters $G$ and regularisation parameter $\lambda$ using our consensus score for consensus COSA clustering yields the highest median ARI at 0.94 (Figure \ref{fig:weighting}A, Supplementary Table 4). The COSA algorithm does not inherently perform feature selection, but we evaluate here its ability to give larger weights to the contributing features using the $F_1$-score measuring the selection performance when considering the $q^* = 20$ features with the largest median weights as selected. The median $F_1$-score is very close to $1$, suggesting that the $q^* = 20$ features with larger median weights are almost always the ones used in the simulation model (Figure \ref{fig:weighting}B). \\

When using the sparse hierarchical clustering algorithm implemented in sparcl, the best clustering performances are as good as those obtained with COSA with the true number of clusters $G^*$ (highest median ARI of $0.95$). However, these correspond to models that are not sparse, with $q = 94$ selected attributes on average (Figure \ref{fig:weighting}A). For sparcl, the $F_1$-score is calculated by considering the $q^* = 20$ features with largest selection proportions as selected. The median $F_1$-score remains below $0.75$ for all values of $\lambda$, suggesting a poorer ability of the sparcl model compared to COSA to give larger weights to the contributing features (Figure \ref{fig:weighting}B, Supplementary Figure 9). Calibration of consensus sparcl clustering using our consensus score yields poor clustering performances, with a median ARI of $0.37$ and a median calibrated number of clusters $\hat{G} = 3$ (Supplementary Table 4). \\

The poor weighting and clustering performances when using sparcl in the proposed approach are likely due to the underlying assumption in sparse clustering that the same set of features equally contribute to the definition of all clusters. As this is, in general, not the case with our simulated data (Supplementary Figures 1, 3), sparcl does not seem to be able to detect the relevant attributes. Supplementary Figure 9 suggests that only a subset of the clusters, supported by a subset of the contributing attributes (in red), are stable in consensus sparcl clustering. These results are in line with previously reported limitations of the sparcl algorithm \cite{rCOSA}. By estimating weights that are specific to the item and attribute, the COSA algorithm generates a distance matrix from which clusters that are supported by different sets of attributes can be detected (Supplementary Figure 9). Based on these results, we recommend the use of the COSA algorithm in the proposed consensus weighted clustering calibrated by maximising the consensus score. \\

Outputs generated by consensus COSA clustering include the estimated cluster membership and the distribution of median feature weights estimated from the $K$ COSA models with calibrated regularisation parameter $\lambda$ (Supplementary Figure 9). We observe increasing median weights with the proportion of explained variance by feature, which suggests that median weights appropriately capture attribute contribution (Supplementary Figure 10). However, the overlapping quartiles of median weights for features with proportions of explained variance below 0.4 indicate that features with weaker contributions may be difficult to disentangle from non-contributing features (Supplementary Figure 10). Introducing correlation between features does not seem to hamper the weighting performances (Supplementary Figure 11). 

\subsection{Real data application}

\subsubsection{Data overview}

We use publicly available data with transcriptomics measurements ($p = 3,312$ attributes) in lung cells from $n=63$ participants \cite{LungCancerSubtypes}. The samples consist of $17$ normal lung specimens and $46$ lung tumours, including histologically defined squamous-cell lung carcinoma ($N=20$), pulmonary carcinoids ($N=20$) and small-cell lung carcinoma ($N=6$). We perform clustering to detect groups of individuals based on their molecular profiles.

\subsubsection{Clustering results}

\begin{figure}[h!]
\centering
\makebox{\includegraphics[width=\linewidth]{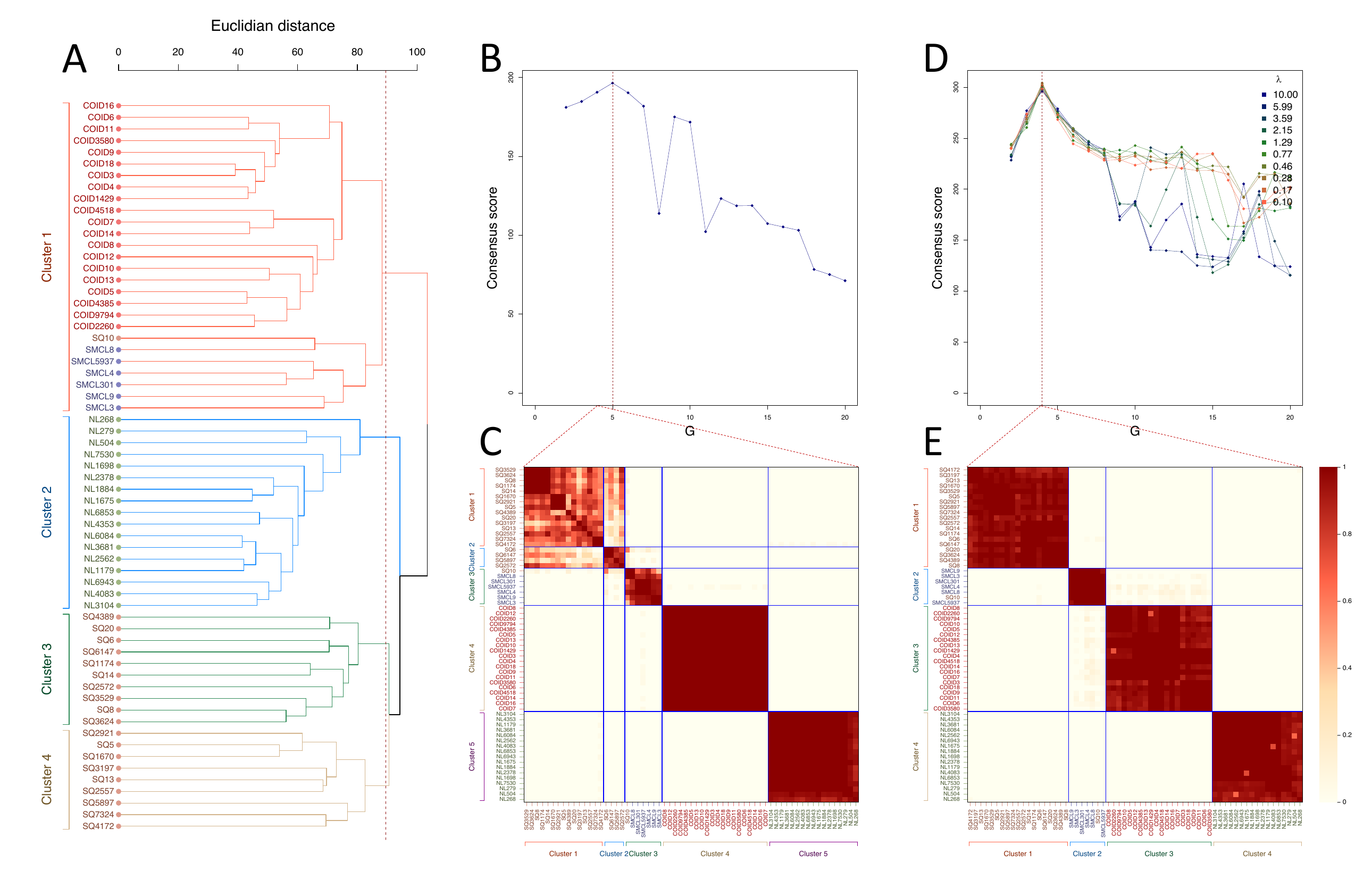}}
\caption{Clustering applications on real transcriptomics data measured in lung tissue. Hierarchical clustering with complete linkage is applied on the Euclidean distance (A). The calibration curve (B) and consensus matrix (C) of consensus clustering using the Euclidean distance are represented. For consensus COSA clustering, we use a grid of $10$ regularisation parameter in the calibration maximising the consensus score (D) and report the calibrated consensus matrix (E). Sample names are coloured by type (normal lung in green, carcinoids in red, squamous-cell in brown and small-cell in blue). For each of these three clustering approaches, the four estimated clusters are coloured in red, blue, green and beige.}
\label{fig:appli}
\end{figure}

First, we apply hierarchical clustering with complete linkage on the Euclidean distances between the $n=63$ samples (Figure \ref{fig:appli}A). As in the original publication, we observe overall lower distances among samples from healthy lungs or from tumours of the same histological subtype than between these groups. The $G^* = 4$ clusters obtained from hierarchical clustering include (i) a mixture of $N=20$ pulmonary carcinoids, $N=6$ small-cell carcinomas and $N=1$ squamous-cell carcinoma in Cluster 1, (i) all normal lung samples in Cluster 2, and (iii) the squamous-cell carcinomas split over Clusters 3 and 4. \\

For consensus unweighted clustering, calibration maximising our score indicates that the most stable clustering is obtained for $\hat{G} = 5$ (Figure \ref{fig:appli}B). The five stable clusters correspond to (i) $N=19$ squamous-cell carcinomas split over Clusters 1 and 2, (ii) all small-cell and the remaining squamous-cell carcinoma in Cluster 3, (iii) all carcinoids in Cluster 4, and (iv) all normal lung tissues in Cluster 5 (Figure \ref{fig:appli}C).\\

For consensus weighted clustering, the calibrated number of clusters is $\hat{G} = 4$ and COSA regularisation parameter is $\hat{\lambda} = 0.28$ (Figure \ref{fig:appli}D). The use of weighted distances induces an increase in the consensus score (highest score of $304$ compared to $196$ in consensus unweighted clustering) that is reflected in the consensus matrix (Figure \ref{fig:appli}E). The $\hat{G} = 4$ stable clusters in the calibrated consensus weighted clustering correspond to the normal lung samples and the three subtypes of lung cancer, except for one squamous-cell sample that is grouped with the small-cell carcinomas in Cluster 2.

\section{Discussion}

As previously reported \cite{MPJ}, we observe better clustering performances with consensus clustering compared to a single run of the underlying (e.g. hierarchical) clustering on both simulated and real data. The use of weighted distances further increases clustering performances in the presence of irrelevant attributes. \\

Our simulation study shows that consensus clustering calibrated by maximising our consensus score performs at least as well as models calibrated using existing scores. Calibration techniques based on cumulative density distributions (PAC and PINS discrepancy scores) only perform well when the clustering structure is extremely strong. Calibration procedures using the consensus and RCSI scores yield similar clustering performances for well separated clusters, but our consensus score allows for the detection of more subtle clustering structures. Furthermore, our consensus score can is far less computational expensive than RCSI scores, which require the simulation and clustering of multiple datasets. \\

The assumption that co-membership probabilities are the same for all pairs within or between consensus clusters, respectively, constitutes a potential limitation of our consensus score. This assumption implies that the quantities $X_w(\lambda, G)$ and $X_b(\lambda, G)$ follow binomial distributions, which allows for the use of the z test and hence for extremely fast computations. As an alternative, the exact distributions could be recovered using simulations. This alternative may quickly become time consuming due to the resolution required to compare the scores. Despite its underlying assumptions and approximations, the simulation studies indicate very good performances of our consensus score. \\

The use of the sparcl or COSA algorithms in consensus weighting clustering both generate an increase in clustering performance compared to unweighted approaches, for well chosen regularisation parameters. Joint calibration of the number of clusters and regularisation parameter in consensus COSA clustering using our consensus score generates some of the best performances. However, the use of sparcl in consensus weighted clustering calibrated using our consensus score is not recommended as it only detects a subset of the clusters, which leads to overall poor clustering performance. Results from our simulation studies are supported by our real data application, where lung sample types are better recovered with consensus weighted clustering calibrated using the consensus score. \\

Due to the subsampling procedure, the increase in performance with consensus clustering comes at the price of a higher computational burden. Our implementation in the R package \texttt{sharp} allows for parallelisation over the subsampling iterations to reduce computation times. The current implementation of consensus clustering does not scale to large numbers of items (e.g. $> 50,000$), which increase both the computation time and memory usage. Extensions potentially using a pre-clustering step are required \cite{FindingGroups, BIRCH}. \\


\section{Data availability}

Consensus (weighted) clustering and proposed calibration procedure have been implemented in the R package sharp (version $\geq$ 1.4.0). Simulation models have been implemented in the R package fake (version $\geq$ 1.4.0). The transcriptomics dataset is publicly available and can be downloaded at \url{https://doi.org/10.1073/pnas.191502998}. All codes to reproduce the analyses presented in this paper are available at \url{https://github.com/barbarabodinier/Consensus_clustering}. 

\section*{Acknowledgements}

We would like to thank Mr Thomas Wright and Mr Ruben Colindres Zuehlke for their insightful suggestions.

\section*{Funding}

This work was supported by the H2020-EXPANSE (Horizon 2020 grant No 874627) and H2020-LongITools (Horizon 2020 grant No 874739) projects. BB received a PhD studentship from the MRC Centre for Environment and Health. 

\section*{Conflict of interest}

MC-H holds shares in the O-SMOSE company and has no conflict of interest to disclose. Consulting activities conducted by the company are independent of the present work. The authors have no conflict of interest to disclose. 

\pagebreak
\bibliography{biblio}
\bibliographystyle{unsrt}

\pagebreak
\section{Supplementary materials}

\subsection{Maximum of the consensus score}

Recall that the integers $X_w (\lambda, G)$ and $X_b (\lambda, G)$ are the total numbers of co-members in the \textit{within} and \textit{between} pairs, respectively. The integers $N_w (\lambda, G)$ and $N_b (\lambda, G)$ are the total numbers of times each of the \textit{within} and \textit{between} pairs, respectively, are drawn together in the subsamples. For clarity, the $\lambda$ and $G$ indexing is omitted here. The consensus score $S_c$ can be expressed as a function of $X_w$, $X_b$, $N_w$ and $N_b$:
\begin{equation}
    \begin{split}
        S_c & = \frac{\frac{X_w}{N_w} - \frac{X_b}{N_b}}{\sqrt{ \left( \frac{X_w+X_b}{N_w+N_b} \right) \left( 1 - \frac{X_w+X_b}{N_w+N_b} \right) \left( \frac{1}{N_w} \frac{1}{N_b} \right)}} \\
        & = \sqrt{N_w + N_b} \frac{\left( \frac{X_w}{N_w} - \frac{X_b}{N_b} \right) \sqrt{N_w N_b}}{\sqrt{ \left( X_w + X_b \right) \left( N_w + N_b - X_w - X_b \right)}}
    \end{split}
\label{eq:z_stat}
\end{equation}

where $N_w \in \mathbb{N}$, $N_b \in \mathbb{N}$, $X_w \in \{ 0, \dots, N_w \}$, $X_b \in \{ 0, \dots, N_b \}$, and $(X_w + X_b) \in \{ 1, \dots, N_w + N_b - 1 \}$. \\

In this section, we want to find the maximum value of the consensus score $S_c$. For this, we introduce the real-valued function $f$ of $X_w$ and $X_b$ which is equal to the consensus score when $X_w$, $X_b$, $N_w$ and $N_b$ are integers:
\begin{equation}
    f(X_w, X_b) = \sqrt{N_w + N_b} \frac{\left( \frac{X_w}{N_w} - \frac{X_b}{N_b} \right) \sqrt{N_w N_b}}{\sqrt{ \left( X_w + X_b \right) \left( N_w + N_b - X_w - X_b \right)}}
\label{eq:f}
\end{equation}
where $N_w > 0$, $N_b > 0$, $X_w \in [0, N_w]$, $X_b \in [0, N_b]$ and $(X_w + X_b) \in [ 1, \dots, N_w + N_b - 1 ]$. \\

We consider that $X_w$ and $X_b$ belong to these intervals in the remainder of this proof. \\

The derivatives of $f$ are given by:
\begin{equation*}
    \small
    \begin{split}
        \frac{\partial f}{\partial X_w} (X_w, X_b) 
        & = \frac{\sqrt{N_w N_b}}{\sqrt{(X_w + X_b) (N_w + N_b - X_w - X_b)}} 
        \left[ \frac{(N_w + N_b) (X_w + 2 X_b + X_b \frac{N_w}{N_b})
        - 2 (X_w + X_b) (X_b + X_b \frac{N_w}{N_b})}{2 N_w (X_w + X_b) (N_w + N_b - X_w - X_b)} \right] \\
        \frac{\partial f}{\partial X_b} (X_w, X_b)
        & = - \frac{\sqrt{N_w N_b}}{\sqrt{(X_w + X_b) (N_w + N_b - X_w - X_b)}} 
        \left[ \frac{(N_w + N_b) (X_b + 2 X_w + X_w \frac{N_b}{N_w})
        - 2 (X_w + X_b) (X_w + X_w \frac{N_b}{N_w})}{2 N_b (X_w + X_b) (N_w + N_b - X_w - X_b)} \right]
    \end{split}
\end{equation*} \\

As $N_w > 0$, $N_b > 0$ and $(X_w + X_b) \in [ 1, \dots, N_w + N_b - 1 ]$, we have
\begin{equation}
    \frac{\sqrt{N_w N_b}}{\sqrt{(X_w + X_b) (N_w + N_b - X_w - X_b)}} \geq 0
    \label{eq:left_term}
\end{equation}
and
\begin{equation}
    2 N_w (X_w + X_b) (N_w + N_b - X_w - X_b) \geq 0
    \label{eq:denom_w}
\end{equation}
as well as 
\begin{equation}
    2 N_b (X_w + X_b) (N_w + N_b - X_w - X_b) \geq 0.
    \label{eq:denom_b}
\end{equation} \\

In addition, we can show that:

\begin{itemize}
    \item If $X_w \geq X_b \frac{N_w}{N_b}$,
\end{itemize}

\begin{equation}
    \begin{split}
        (N_w + N_b) (X_w + 2 X_b + X_b \frac{N_w}{N_b}) 
        - 2 \underbrace{(X_w + X_b)}_{< (N_w + N_b)} (X_b + X_b \frac{N_w}{N_b}) \\
        > (N_w + N_b) (X_w + 2 X_b + X_b \frac{N_w}{N_b}) 
        - 2 (N_w + N_b) (X_b + X_b \frac{N_w}{N_b}) \\
        = (N_w + N_b) (X_w + 2 X_b + X_b \frac{N_w}{N_b} - 2 X_b - 2 X_b \frac{N_w}{N_b}) \\
        = (N_w + N_b) (X_w - X_b \frac{N_w}{N_b}) \\
        \geq 0
    \end{split}
    \label{eq:num_w_1}
\end{equation}
as $(X_w + X_b )\leq (N_w + N_b - 1)$. 

\begin{itemize}
    \item If $X_w \leq X_b \frac{N_w}{N_b}$,
\end{itemize}

\begin{equation}
    \begin{split}
        (N_w + N_b) (X_w + 2 X_b + X_b \frac{N_w}{N_b}) 
        - 2 (X_w + X_b) (X_b + X_b \frac{N_w}{N_b}) \\
        \geq (N_w + N_b) (2 X_w + 2 X_b) 
        - 2 (X_w + X_b) (X_b + X_b \frac{N_w}{N_b}) \\
        = 2 (X_w + X_b) (N_w + N_b - \underbrace{X_b}_{\leq N_b} - \underbrace{X_b}_{\leq N_b} \frac{N_w}{N_b}) \\
        \geq 2 (X_w + X_b) (N_w + N_b - N_b - N_b \frac{N_w}{N_b}) \\
        = 0
    \end{split}
    \label{eq:num_w_2}
\end{equation}
as $X_b \leq N_b$. 

Equations \ref{eq:num_w_1} and \ref{eq:num_w_2} show that 
\begin{equation}
    (N_w + N_b) (X_w + 2 X_b + X_b \frac{N_w}{N_b}) 
        - 2 (X_w + X_b) (X_b + X_b \frac{N_w}{N_b}) \geq 0
    \label{eq:num_w}
\end{equation}
for any values of $X_w$ and $X_b$ over the intervals where $f$ is defined. \\

Similarly, we can show that 
\begin{equation}
    (N_w + N_b) (X_b + 2 X_w + X_w \frac{N_b}{N_w})
        - 2 (X_w + X_b) (X_w + X_w \frac{N_b}{N_w}) \geq 0.
    \label{eq:num_b}
\end{equation}

Combining Equations \ref{eq:left_term}, \ref{eq:denom_w} and \ref{eq:num_w}, we can show that $f$ is monotonically non-decreasing over $X_w$ as
\begin{equation*}
    \frac{\partial f}{\partial X_w} (X_w, X_b) \geq 0.
\end{equation*}

Combining Equations \ref{eq:left_term}, \ref{eq:denom_b} and \ref{eq:num_b}, we can show that $f$ is monotonically non-increasing over $X_b$ as
\begin{equation*}
    \frac{\partial f}{\partial X_b} (X_w, X_b) \leq 0.
\end{equation*}

Hence, the function $f$ is maximised at $X_w = N_w$ and $X_b = 0$, which are the largest and smallest values for $X_w$ and $X_b$, respectively. The corresponding maximum is:
\begin{equation*}
    f(X_w = N_w, X_b = N_b) = \sqrt{N_w + N_b}
\end{equation*}

The consensus score can be obtained by applying the function $f$ on integers defined over the same intervals (see Equation \ref{eq:z_stat}). As a consequence, the consensus score is also maximised at $X_w = N_w$ and $X_b = 0$, which corresponds to a binary consensus matrix. \\

To illustrate this result, we represent the values of the consensus score $S_c$ obtained with different values of $X_w$ (x-axis) and $X_b$ (y-axis) in the heatmap below (Figure A). In this example, we used $N_w = 10$ and $N_b = 20$. 

\pagebreak

\begin{figure}[h!]
\centering
\makebox{\includegraphics[width=0.7\linewidth]{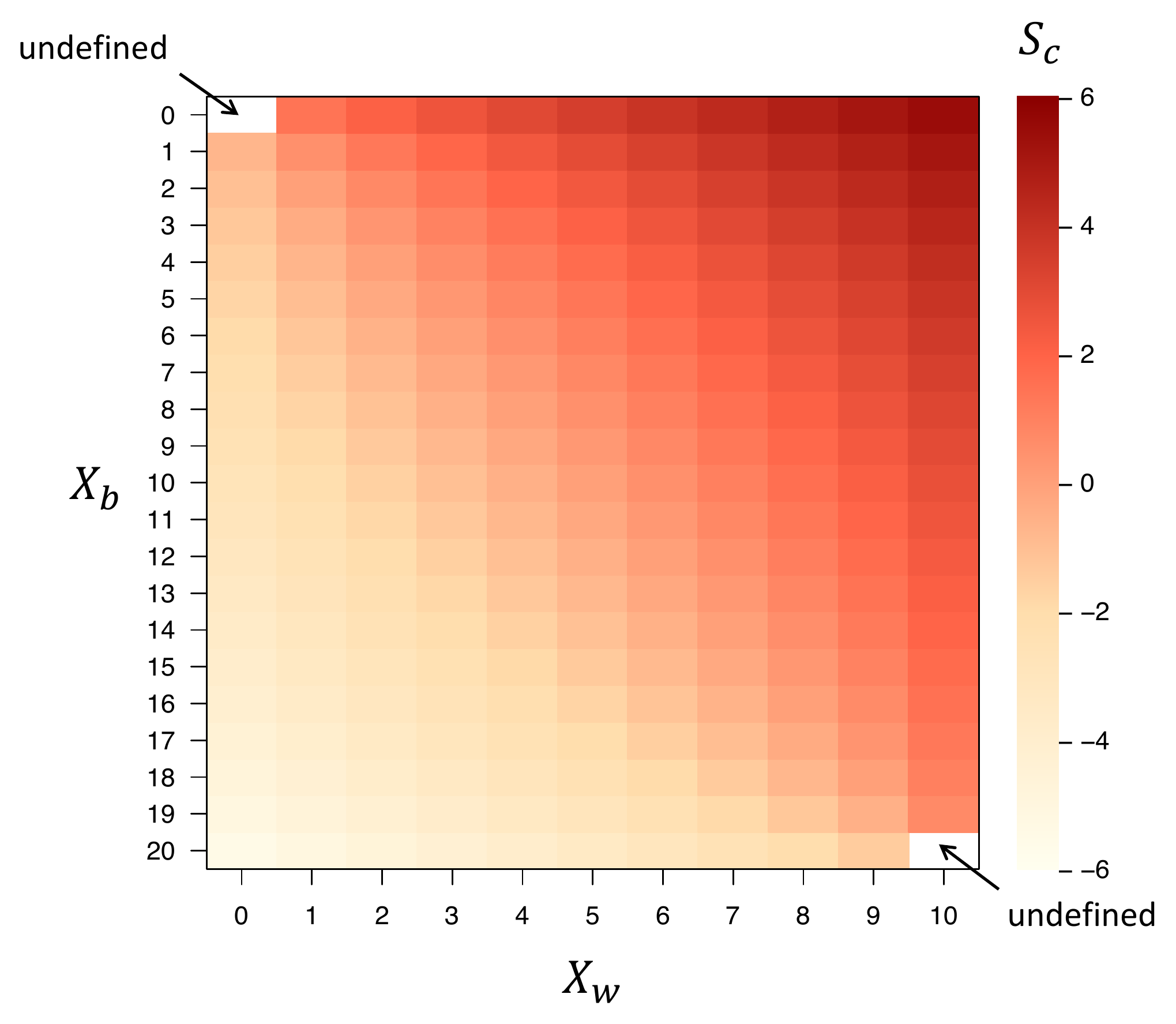}}
\end{figure}
Figure A: Heatmap of the consensus score (colour-coded) obtained with $N_w = 10$ and $N_b = 20$ and different values of $X_w$ (x-axis) and $X_b$ (y-axis). 

\pagebreak

\begin{figure}[h!]
\centering
\makebox{\includegraphics[width=\linewidth]{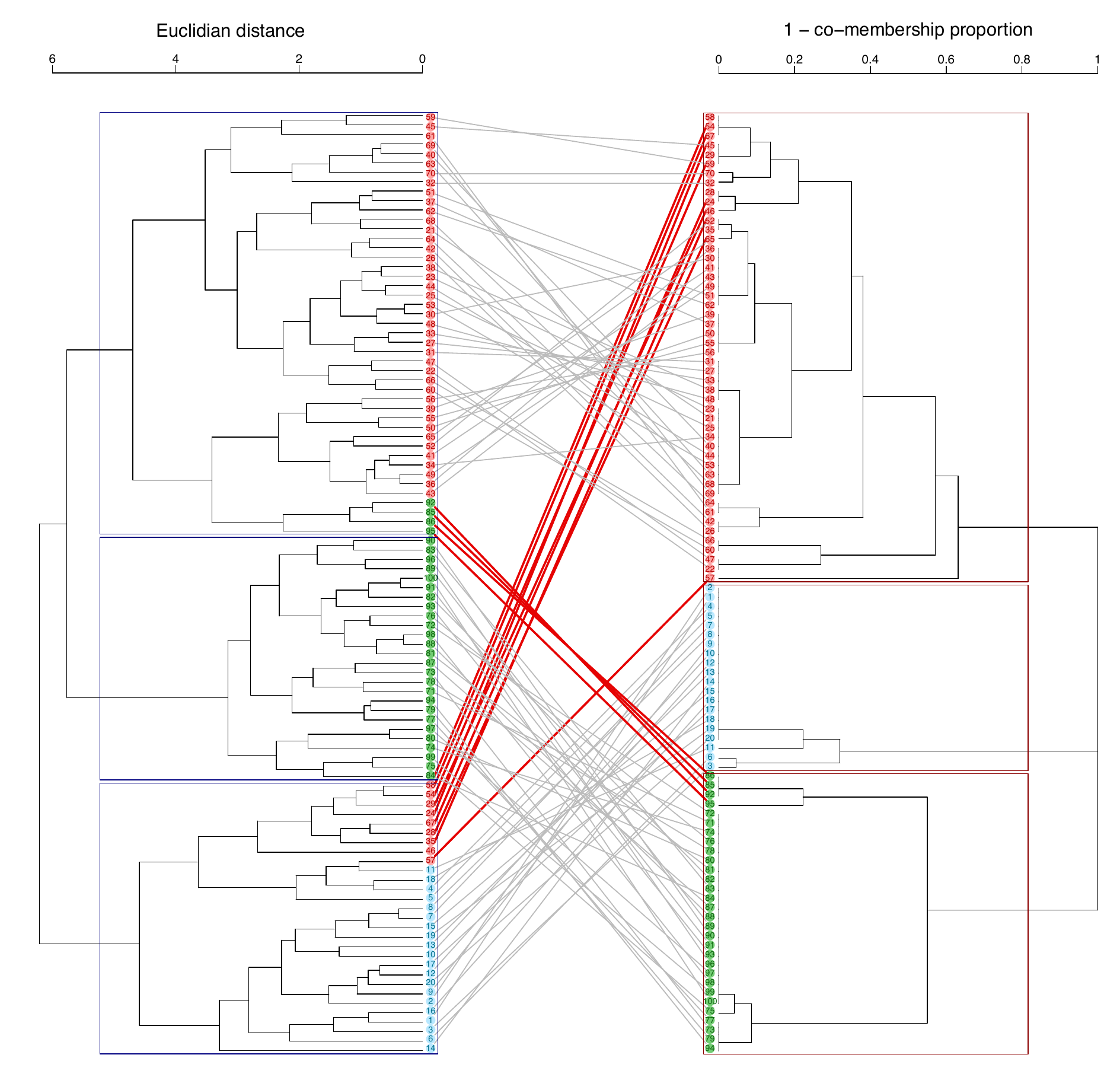}}
\end{figure}
Supplementary Figure 1: Comparison of cluster membership by applying hierarchical clustering with complete linkage using the Euclidean distance (left) or 1 - co-membership proportion (right) as a distance measure. The co-membership proportions are obtained from hierarchical clustering on the Euclidean distances calculated on $K=100$ subsamples. We represent the dendrograms obtained with the two approaches. The items are coloured by true (simulated) cluster membership. The same items are present in both dendrograms but may be re-ordered due to the change of distance metric. Positions of the same item in the two dendrograms are connected by edges to improve readability. Cluster re-assignment is indicated by red edges.  
\pagebreak

\begin{figure}[h!]
\centering
\makebox{\includegraphics[width=\linewidth]{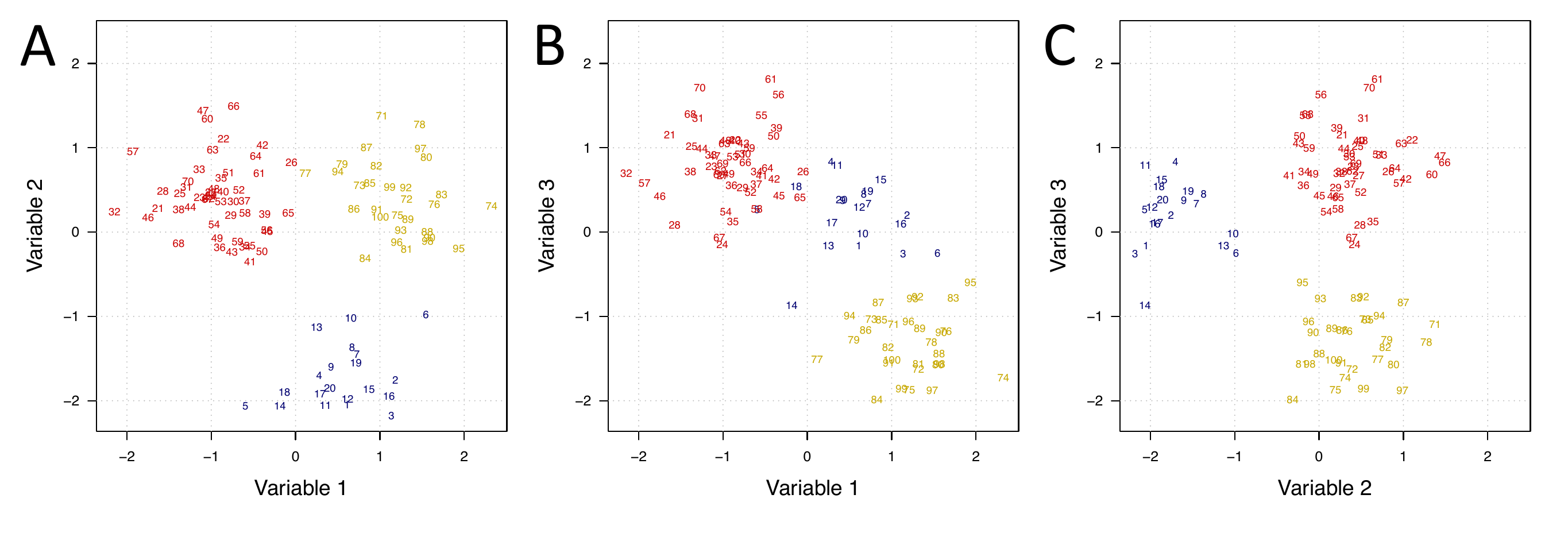}}
\end{figure}
Supplementary Figure 2: Scatter plots for simulated data using $n=100$ items split into $G^* = 3$ clusters such that $N_1 = 20$ (in blue), $N_2 = 50$ (in red) and $N_3 = 30$ (in orange) across $p=3$ features with a proportion of explained variance by the grouping structure set to $E=0.8$ for all features. 
\pagebreak

\begin{figure}[h!]
\centering
\makebox{\includegraphics[width=\linewidth]{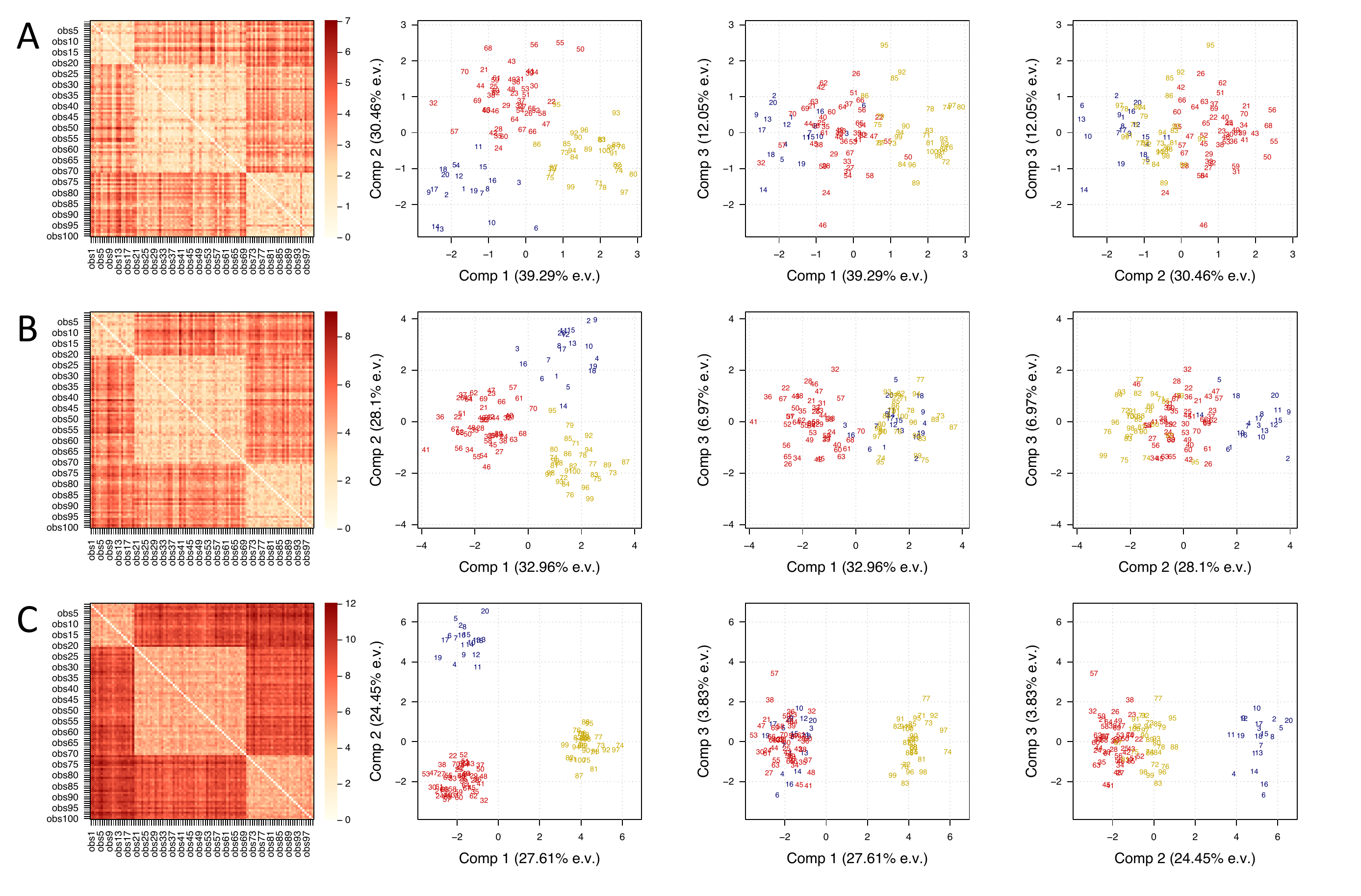}}
\end{figure}
Supplementary Figure 3: Example of simulated data using $n=100$ items split into $G^* = 3$ clusters such that $N_1 = 20$ (in blue), $N_2 = 50$ (in red) and $N_3 = 30$ (in orange) across $p=5$ (A), $p=10$ (B) or $p=30$ (C) features with a proportion of explained variance by the grouping structure set to $E=0.5$ for all features. For each dataset, we show the heatmap of Euclidean distances (left) and score plots along the first three principal components of a Principal Component Analysis (right). 
\pagebreak

\begin{figure}[h!]
\centering
\makebox{\includegraphics[width=\linewidth]{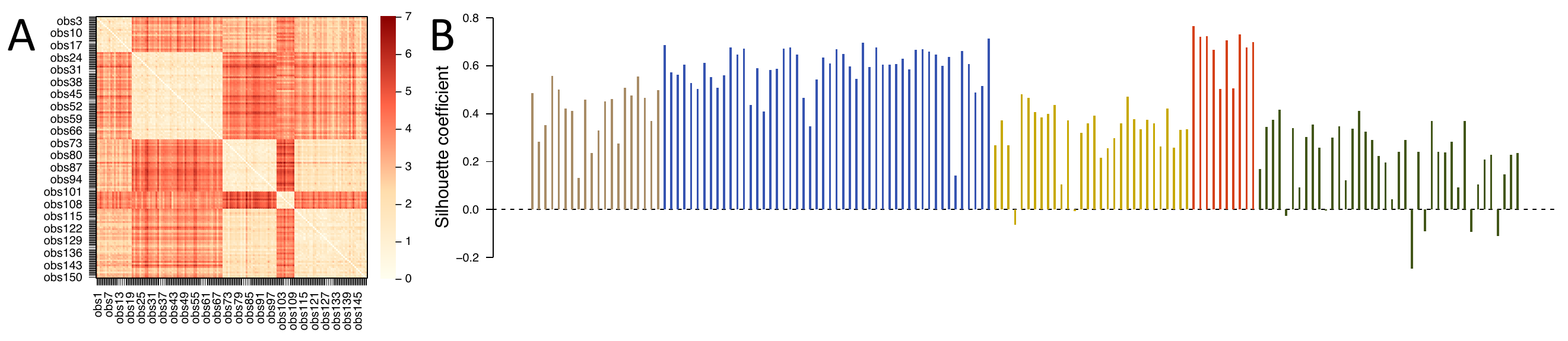}}
\end{figure}
Supplementary Figure 4: Example of simulated data using $n=150$ items split into $G^* = 5$ clusters such that $N_1 = 20$, $N_2 = 50$, $N_3 = 30$, $N_4 = 10$, $N_5 = 40$ across $p=5$ features with a proportion of explained variance by the grouping structure set to $E=0.8$ for all features. We show the heatmap of Euclidean distances (A) and silhouette widths for each of the $n = 150$ items coloured by simulated cluster membership (B). 
\pagebreak

\begin{figure}[h!]
\centering
\makebox{\includegraphics[width=0.7\linewidth]{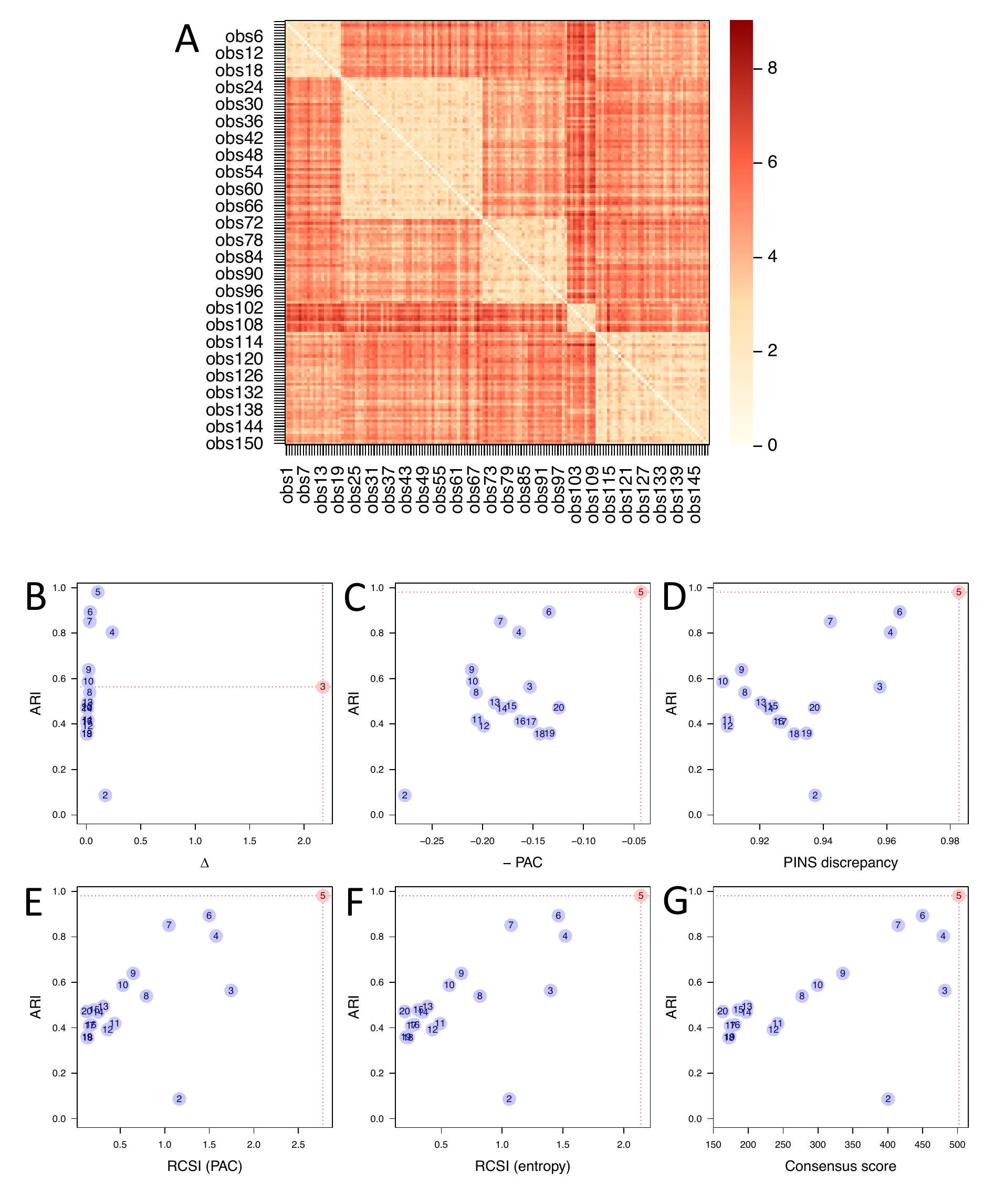}}
\end{figure}
Supplementary Figure 5: Consensus unweighted clustering performance as a function of different calibration scores applied on simulated data with $G^* = 5$ clusters. Consensus clustering was conducted using hierarchical clustering with complete linkage on the Euclidean distances computed on $K=100$ subsamples. We show the heatmap of pairwise distances in the simulated dataset with $n=150$ items split into $G^* = 5$ clusters such that $N_1 = 20$, $N_2 = 50$, $N_3 = 30$, $N_4 = 10$, $N_5 = 40$ across $p=10$ features, each with a proportion of explained variance of $0.6$ (A). The Adjusted Rand Index (ARI) measuring clustering performance is represented as a function of the $\Delta$ (B), PAC (C), PINS discrepancy (D), RCSI for PAC (E), RCSI for entropy (F) and consensus (G) scores for different numbers of clusters (indicated on the points). The calibrated number of clusters obtained with the corresponding calibration score is indicated in red.
\pagebreak

\begin{figure}[h!]
\centering
\makebox{\includegraphics[width=0.87\linewidth]{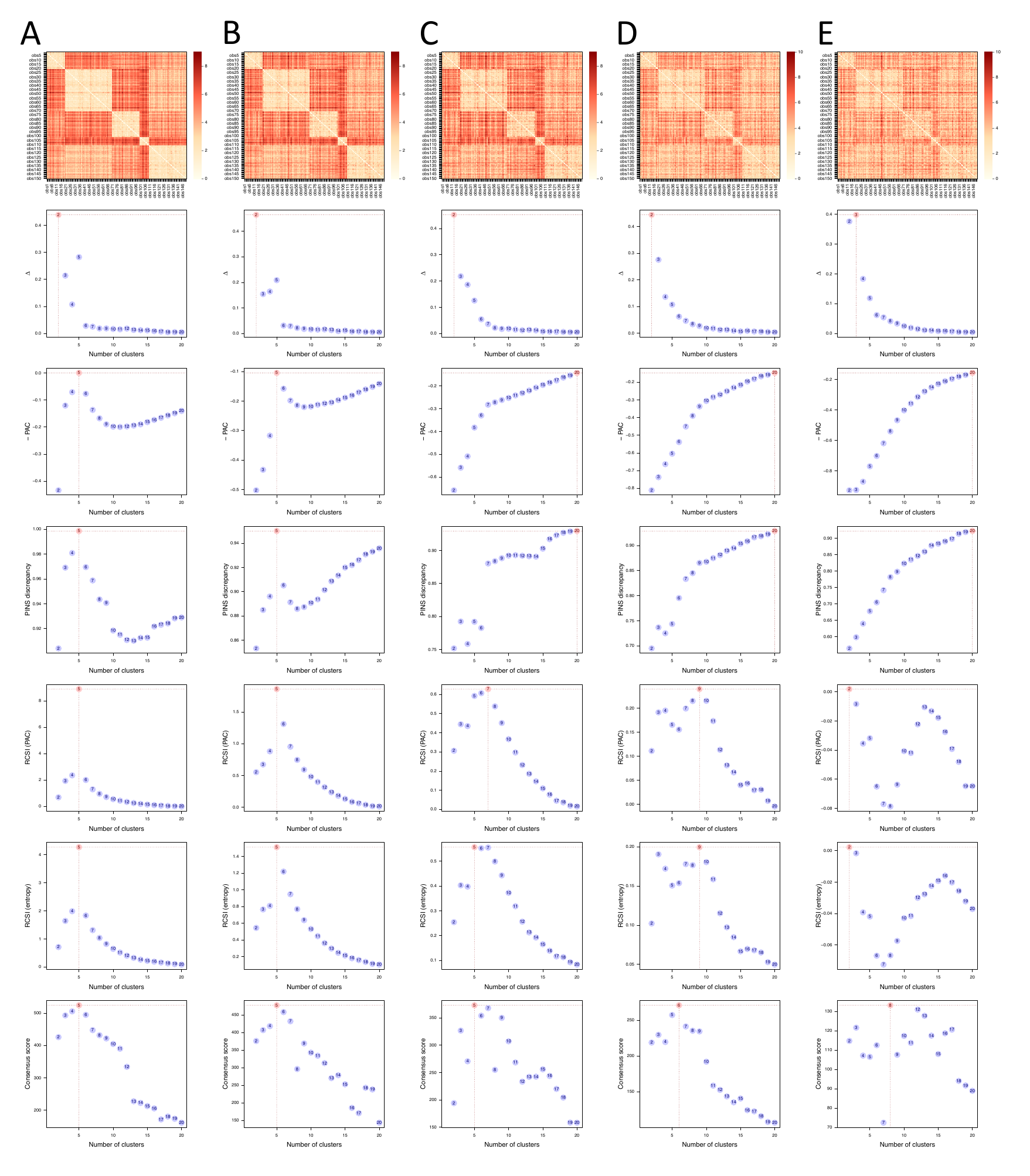}}
\end{figure}
Supplementary Figure 6: Calibration curves using the PAC, $\Delta$, PINS discrepancy, RCSI and consensus scores on simulated examples with different levels of cluster separation and compactness. Consensus clustering was conducted using hierarchical clustering with complete linkage on the Euclidean distances computed on $K=100$ subsamples. Data is simulated for $n=150$ items split into $G^* = 5$ clusters such that $N_1 = 20$, $N_2 = 50$, $N_3 = 30$, $N_4 = 10$, $N_5 = 40$ across $p=10$ features with a proportion of explained variance by the grouping structure set to $E=0.7$ (A), $E=0.6$ (B), $E=0.5$ (C), $E=0.4$ (D), or $E=0.3$ (E) for all features. Heatmaps of Euclidean distances for calculated for the simulated data are reported at the top. 
\pagebreak

\begin{figure}[h!]
\centering
\makebox{\includegraphics[width=\linewidth]{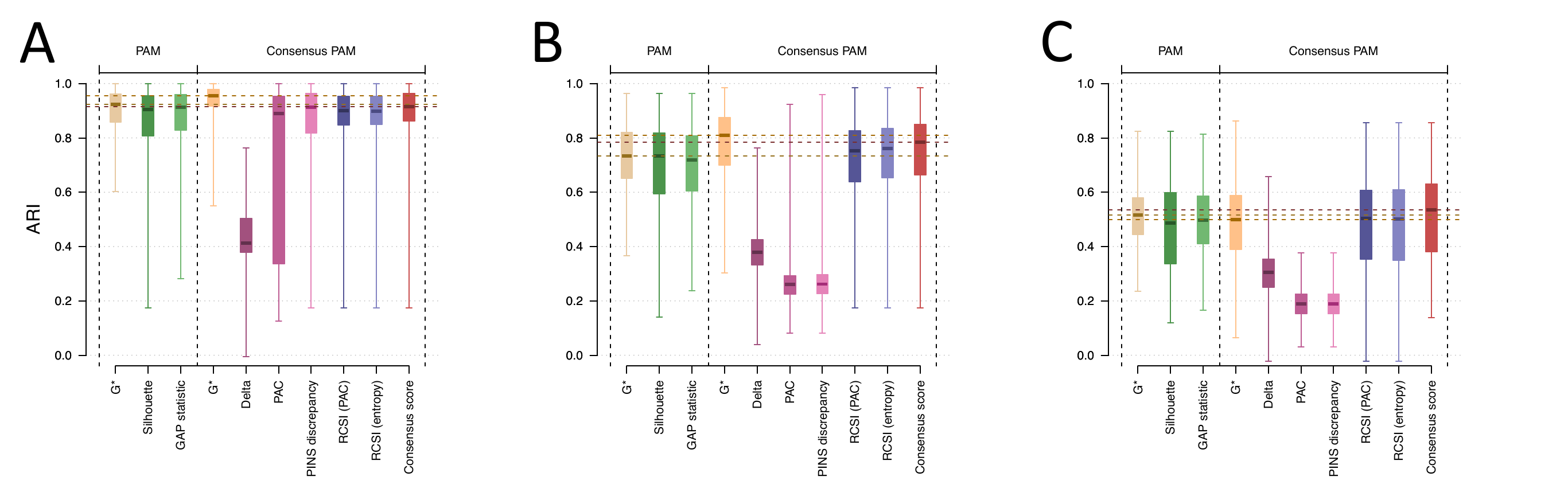}}
\end{figure}
Supplementary Figure 7: Comparison of clustering performances of (consensus) Partitioning Around Medoids (PAM) with different calibration strategies from $N=1,000$ simulated datasets corresponding to different levels of cluster separation. We simulate $N=1,000$ datasets with $n=150$ items split into $G^* = 5$ clusters such that $N_1 = 20$, $N_2 = 50$, $N_3 = 30$, $N_4 = 10$, $N_5 = 40$ across $p=10$ features, each with a proportion of explained variance of $E=0.6$ (left), $E=0.5$ (middle) or $E=0.4$ (right). Median, quartiles, minimum and maximum Adjusted Rand Index (ARI) for PAM with the simulated number of clusters ($G^*$), or calibrated by maximising the silhouette and GAP score, and for consensus PAM with $G^*$ or calibrated using the $\Delta$, PAC, PINS discrepancy, RCSI and consensus scores are reported.
\pagebreak

\begin{figure}[h!]
\centering
\makebox{\includegraphics[width=\linewidth]{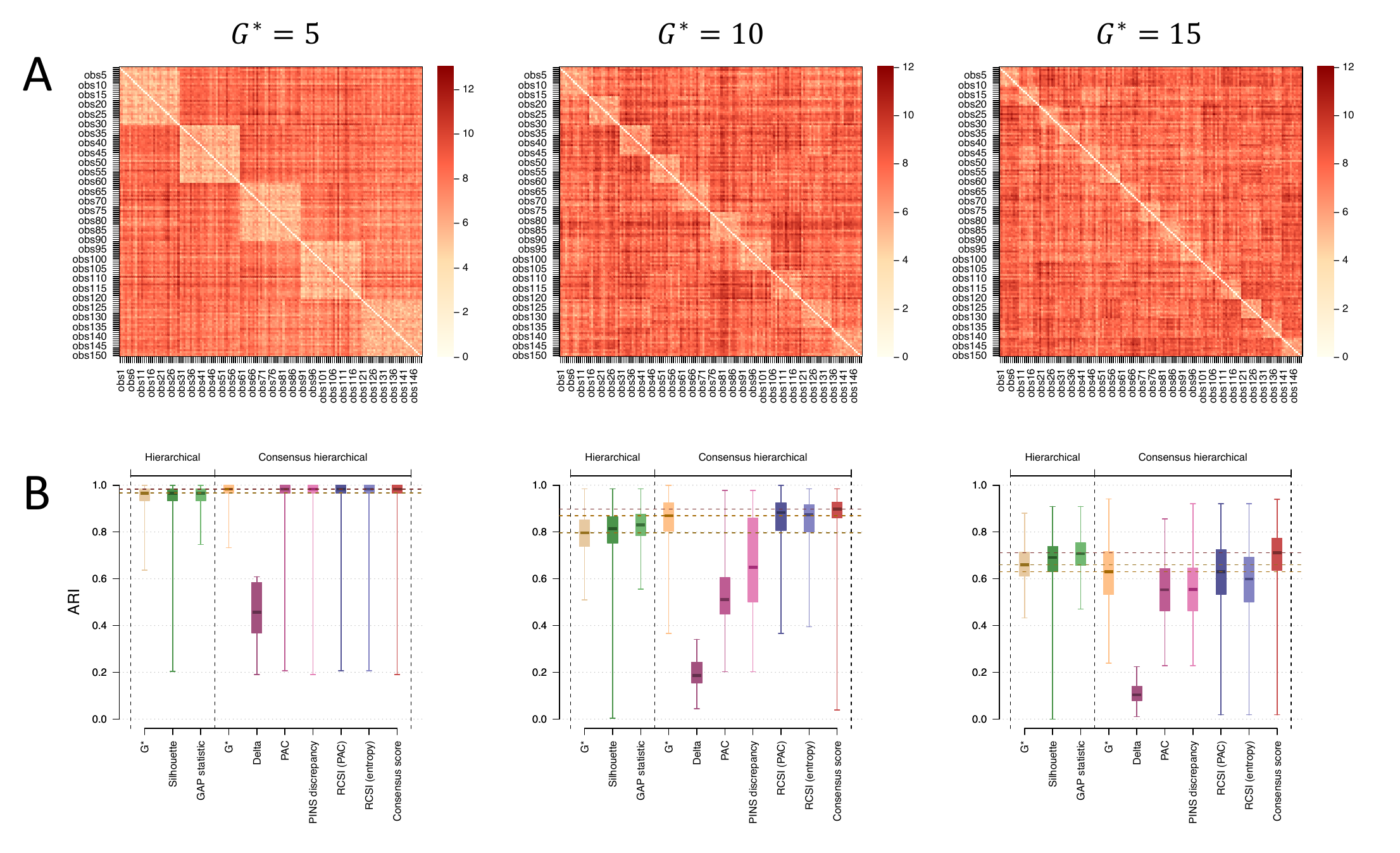}}
\end{figure}
Supplementary Figure 8: Comparison of clustering performances of (consensus) hierarchical clustering with different calibration strategies from $N=1,000$ simulated datasets corresponding to different numbers of clusters. We simulate $N=1,000$ datasets with $n=150$ items split into $G^* = 5$ (left), $G=10$ (middle), or $G=15$ (right) clusters of equal sizes across $p=10$ features, each with a proportion of explained variance of $E=0.5$. For each scenario, we show a heatmap of Euclidean distances (A). Median, quartiles, minimum and maximum Adjusted Rand Index (ARI) for hierarchical clustering with the simulated number of clusters ($G^*$), or calibrated by maximising the silhouette and GAP score, and for consensus hierarchical clustering with $G^*$ or calibrated using the $\Delta$, PAC, PINS discrepancy, RCSI and consensus scores are reported (B).
\pagebreak

\begin{figure}[h!]
\centering
\makebox{\includegraphics[width=\linewidth]{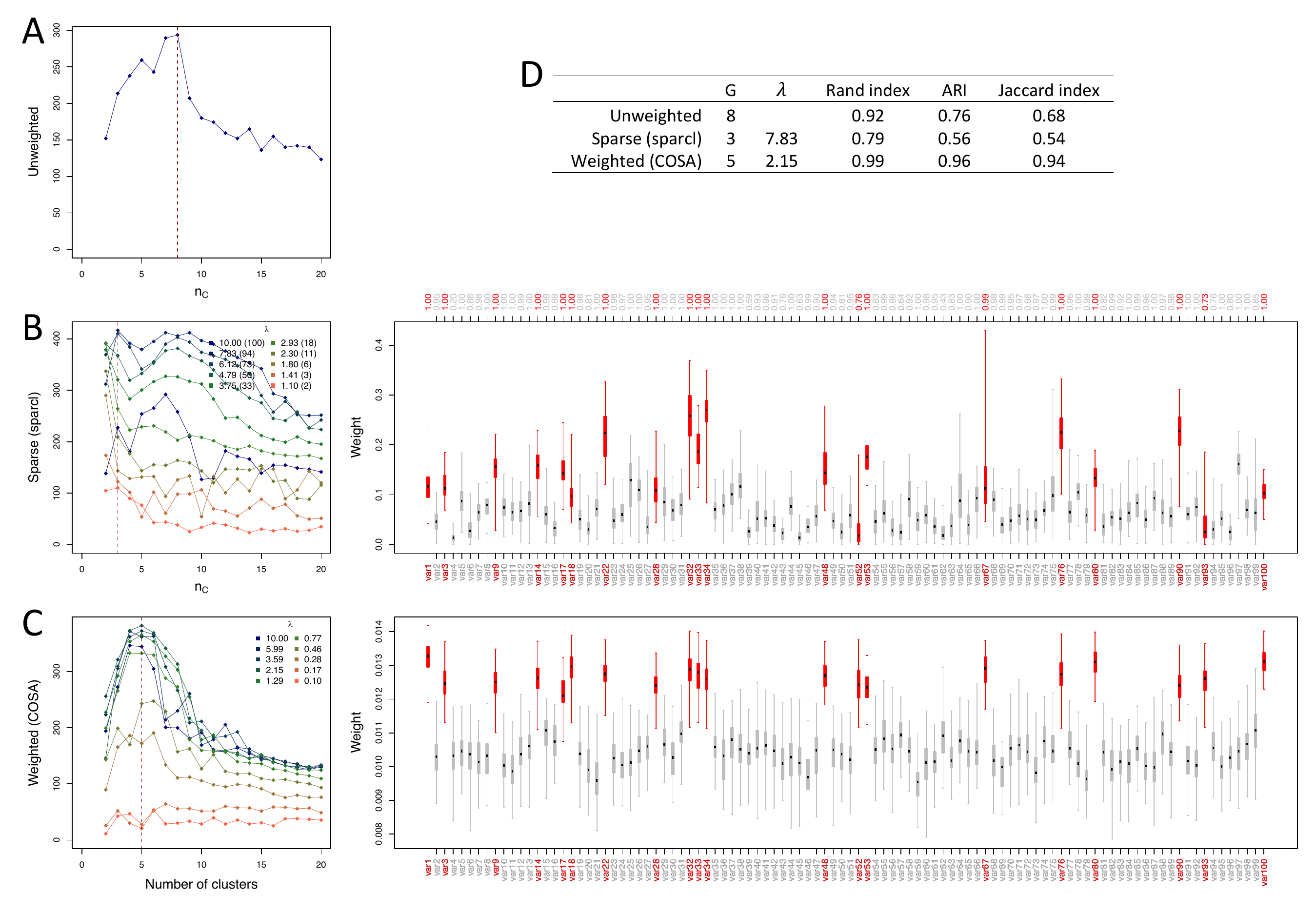}}
\end{figure}
Supplementary Figure 9: Calibration and performances of consensus clustering applied on sparse hierarchical clustering (sparcl) or hierarchical clustering using COSA distances (COSA). The simulated datasets has $n=150$ items split into $G^* = 5$ such that $N_1 = 20$, $N_2 = 50$, $N_3 = 30$, $N_4 = 10$, $N_5 = 40$ across $p=100$ features, of which $q^* = 20$ had a nonzero proportion of explained variance ($E=0.6$). Calibration curves show the consensus score as a function of the number of clusters for consensus unweighted (A), sparcl (B) and COSA (C) clustering. The curves obtained with different $\lambda$ values are showed for sparcl and COSA (B, C). Boxes showing the distribution of (median) feature weights obtained with sparcl and COSA are coloured in red for contributing features ($E=0.6$) and grey for non-contributing features ($E=0$) (B, C). The selection proportions are showed on the top for sparcl (B). Clustering performances are reported for consensus unweighted, sparcl and COSA clustering (D). 
\pagebreak

\begin{figure}[h!]
\centering
\makebox{\includegraphics[width=\linewidth]{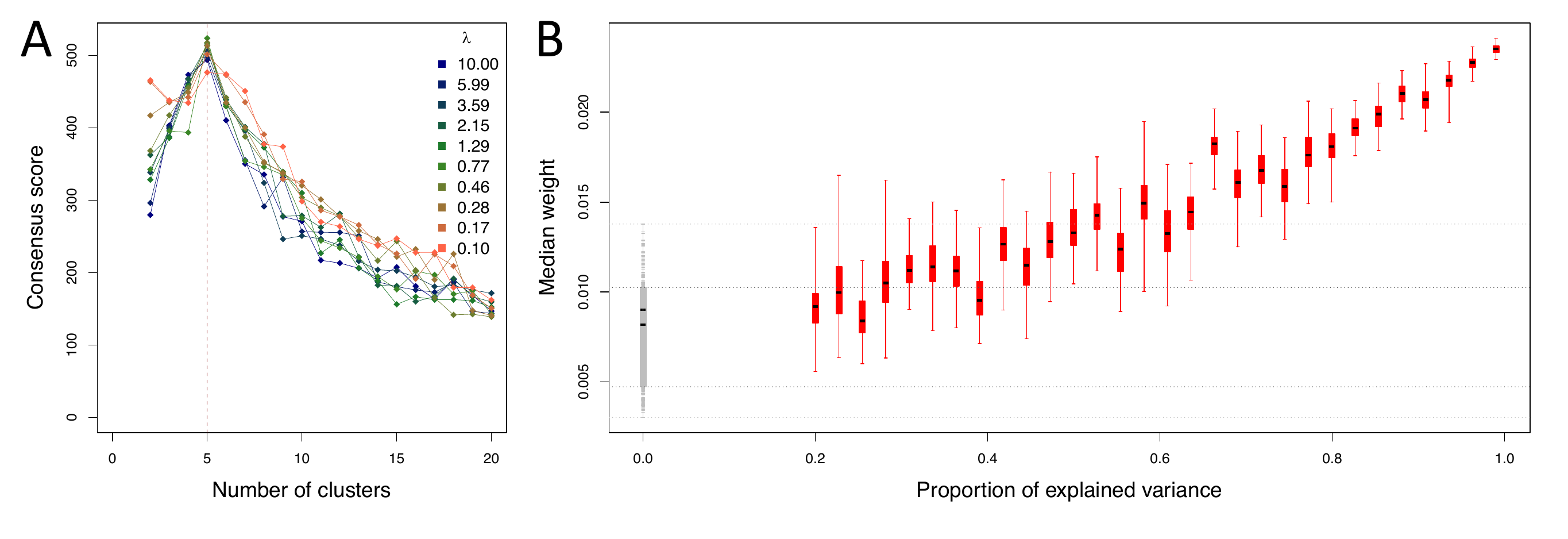}} 
\end{figure}
Supplementary Figure 10: Distribution of median weights obtained from consensus COSA clustering as a function of the simulated proportion of explained variance by feature. The simulated datasets has $n=150$ items split into $G^* = 5$ such that $N_1 = 20$, $N_2 = 50$, $N_3 = 30$, $N_4 = 10$, $N_5 = 40$ across $p=100$ features, of which $q^* = 30$ had a nonzero proportion of explained variance (ranging from $E=0.2$ to $E=0.99$). Calibration curves show the consensus score as a function of the number of clusters (A). Boxes showing the distribution of (median) feature weights are coloured in red for contributing features ($E \neq 0$) and grey for non-contributing features ($E=0$) (B). Features are ordered by proportion of simulated explained variance. 
\pagebreak

\begin{figure}[h!]
\centering
\makebox{\includegraphics[width=\linewidth]{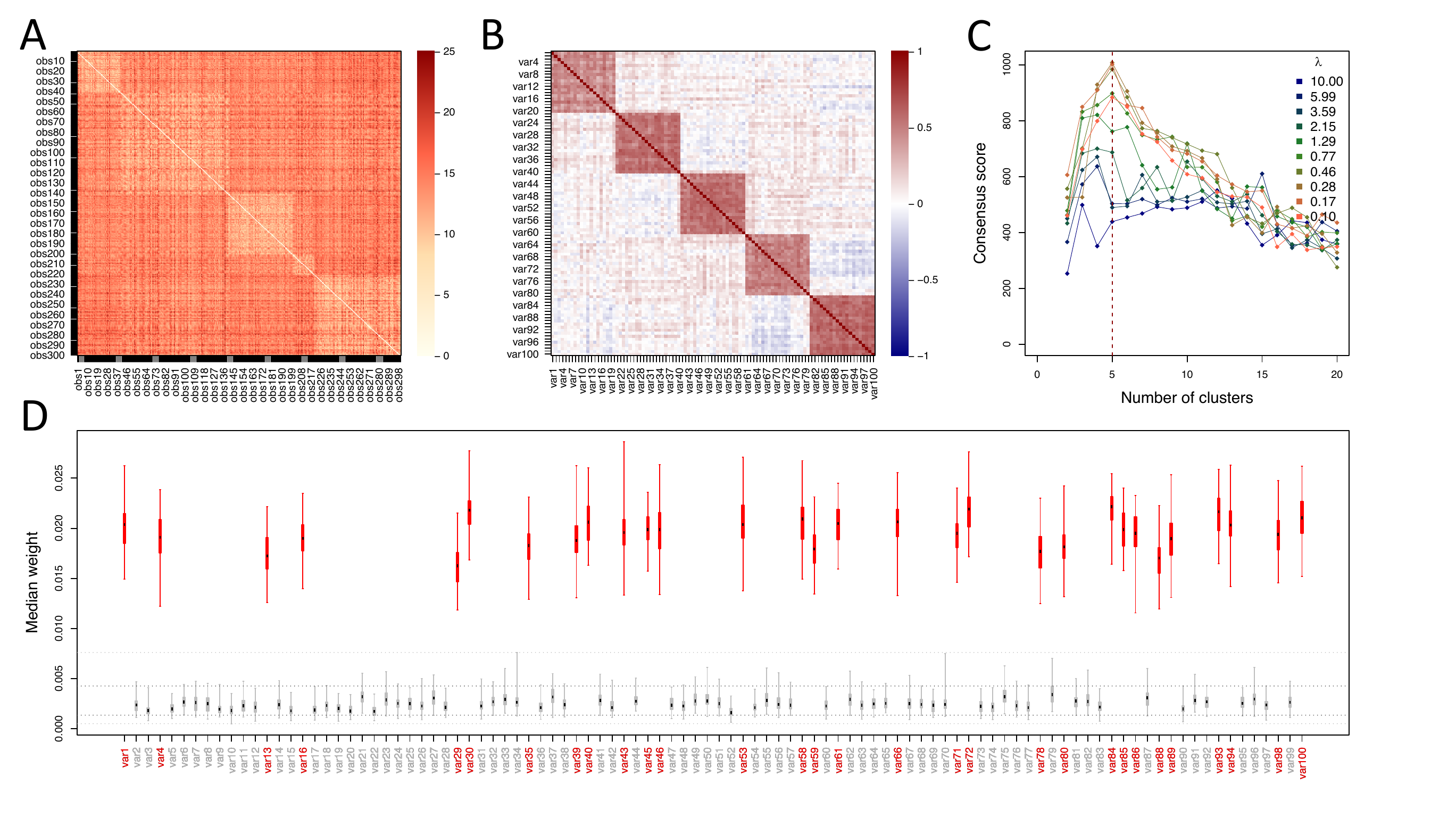}}
\end{figure}
Supplementary Figure 11: Distribution of median weights obtained from consensus COSA clustering on simulated data with correlated features. The simulated datasets has $n=300$ items split into $G^* = 5$ such that $N_1 = 40$, $N_2 = 100$, $N_3 = 60$, $N_4 = 20$, $N_5 = 80$ across $p=100$ features, of which $q^* = 30$ had a nonzero proportion of explained variance ($E=0.7$). We show heatmaps of Euclidean distances between items (A) and Pearson's correlations between features (B). Calibration curves show the consensus score as a function of the number of clusters (C). Boxes showing the distribution of (median) feature weights are coloured in red for contributing features ($E = 0.7$) and grey for non-contributing features ($E=0$) (D). 
\pagebreak

\noindent
\renewcommand{\arraystretch}{0.8}
{\fontsize{10}{10}\selectfont
\begin{tabular}{cccccccc} \hline
E&Method&G&Rand index&ARI&Jaccard index&Percentage&Time (s)\\ \hline
&G*&5 [0]&0.971 [0.029]&0.921 [0.079]&0.887 [0.106]&&0 [0] \\
&Silhouette&5 [1]&0.957 [0.110]&0.884 [0.251]&0.840 [0.275]&&0 [0] \\
&GAP statistic&5 [0]&0.972 [0.027]&0.922 [0.075]&0.887 [0.100]&&1 [0] \\
&G*&5 [0]&0.984 [0.018]&0.957 [0.050]&0.936 [0.071]&&3 [1] \\ \cline{2-8}
0.6&Delta&3 [1]&0.740 [0.127]&0.480 [0.199]&0.480 [0.139]&&3 [1] \\
&PAC&5 [0]&0.978 [0.192]&0.940 [0.493]&0.912 [0.481]&&3 [1] \\
&PINS discrepancy&5 [1]&0.978 [0.107]&0.939 [0.247]&0.911 [0.289]&&3 [1] \\
&RCSI (PAC)&5 [1]&0.979 [0.049]&0.942 [0.126]&0.916 [0.168]&100\%&162 [46] \\
&RCSI (entropy)&5 [1]&0.980 [0.038]&0.945 [0.100]&0.919 [0.135]&100\%&149 [42] \\
&Consensus score&5 [1]&0.979 [0.053]&0.943 [0.134]&0.917 [0.173]&&3 [1] \\ \hline
&G*&5 [0]&0.916 [0.053]&0.770 [0.143]&0.703 [0.156]&&0 [0] \\
&Silhouette&4 [3]&0.866 [0.237]&0.664 [0.403]&0.607 [0.314]&&0 [0] \\
&GAP statistic&5 [0]&0.918 [0.047]&0.776 [0.132]&0.707 [0.147]&&1 [0] \\
&G*&5 [0]&0.948 [0.037]&0.857 [0.100]&0.805 [0.122]&&3 [1] \\ \cline{2-8}
0.5&Delta&3 [1]&0.730 [0.123]&0.456 [0.176]&0.463 [0.116]&&3 [1] \\
&PAC&20 [0]&0.770 [0.032]&0.280 [0.102]&0.267 [0.081]&&3 [1] \\
&PINS discrepancy&20 [0]&0.771 [0.034]&0.283 [0.114]&0.274 [0.092]&&3 [1] \\
&RCSI (PAC)&5 [1]&0.936 [0.083]&0.826 [0.200]&0.767 [0.218]&100\%&157 [42] \\
&RCSI (entropy)&5 [1]&0.939 [0.080]&0.832 [0.195]&0.774 [0.212]&100\%&148 [35] \\
&Consensus score&5 [1]&0.940 [0.085]&0.836 [0.200]&0.779 [0.216]&&3 [1] \\ \hline
&G*&5 [0]&0.831 [0.060]&0.540 [0.152]&0.485 [0.126]&&0 [0] \\
&Silhouette&3 [2]&0.712 [0.262]&0.391 [0.337]&0.408 [0.186]&&0 [0] \\
&GAP statistic&5 [2]&0.840 [0.050]&0.555 [0.144]&0.497 [0.130]&&1 [0] \\
&G*&5 [0]&0.874 [0.064]&0.655 [0.164]&0.586 [0.152]&&2 [1] \\ \cline{2-8}
0.4&Delta&3 [1]&0.707 [0.117]&0.405 [0.163]&0.423 [0.100]&&2 [1] \\
&PAC&20 [0]&0.736 [0.038]&0.197 [0.090]&0.218 [0.065]&&2 [1] \\
&PINS discrepancy&20 [0]&0.736 [0.039]&0.197 [0.091]&0.218 [0.067]&&2 [1] \\
&RCSI (PAC)&5 [2]&0.854 [0.089]&0.608 [0.218]&0.542 [0.184]&100\%&142 [36] \\
&RCSI (entropy)&5 [2]&0.855 [0.092]&0.610 [0.223]&0.544 [0.184]&100\%&133 [37] \\
&Consensus score&5 [2]&0.868 [0.103]&0.642 [0.217]&0.579 [0.181]&&2 [1] \\ \hline
\end{tabular}}
\vspace{10pt}

Supplementary Table 1: Clustering performances of (consensus) hierarchical clustering with different calibration strategies from $N=1,000$ simulated datasets corresponding to different levels of cluster separation. We simulate $N=1,000$ datasets with $n=150$ items split into $G^* = 5$ clusters such that $N_1 = 20$, $N_2 = 50$, $N_3 = 30$, $N_4 = 10$, $N_5 = 40$ across $p=10$ features, each with a proportion of explained variance of $E=0.6$ (top), $E=0.5$ (middle) or $E=0.4$ (bottom). Median and inter-quartile range of the calibrated number of clusters (G), Rand index, Adjusted Rand Index (ARI), Jaccard index, and computation time in seconds for hierarchical clustering with the simulated number of clusters ($G^*$), or calibrated by maximising the silhouette and GAP score, and for consensus hierarchical clustering with $G^*$ or calibrated using the $\Delta$, PAC, PINS discrepancy, RCSI and consensus scores are reported. The RCSI is computed using $N=25$ iterations. We also report the percentage of significant clustering structure for Monte Carlo approaches. For consensus clustering, the reported time includes both the computation of consensus matrices and the calibration procedure. 
\pagebreak

\noindent
\renewcommand{\arraystretch}{0.8}
\begin{center}
{\fontsize{10}{11}\selectfont
\begin{tabular}{cccccccc} \hline
E&Method&G&Rand index&ARI&Jaccard index&Percentage&Time (s)\\ \hline
0.6&RCSI (PAC)&5 [1]&0.979 [0.049]&0.942 [0.128]&0.916 [0.169]&100\%&642 [264]\\
&RCSI (entropy)&5 [1]&0.980 [0.038]&0.945 [0.100]&0.919 [0.135]&100\%&708 [394]\\ \hline
0.5&RCSI (PAC)&5 [1]&0.936 [0.082]&0.824 [0.199]&0.765 [0.214]&100\%&655 [234]\\
&RCSI (entropy)&5 [1]&0.939 [0.080]&0.832 [0.194]&0.774 [0.211]&100\%&718 [298]\\ \hline
0.4&RCSI (PAC)&5 [2]&0.854 [0.090]&0.606 [0.219]&0.541 [0.187]&100\%&808 [355]\\
&RCSI (entropy)&5 [2]&0.855 [0.091]&0.607 [0.222]&0.542 [0.186]&100\%&814 [382]\\ \hline
\end{tabular}}
\end{center}
\vspace{10pt}

Supplementary Table 2: Clustering performances of (consensus) hierarchical clustering calibrated by RCSI score using $N = 100$ iterations from $N=1,000$ simulated datasets corresponding to different levels of cluster separation. We simulate $N=1,000$ datasets with $n=150$ items split into $G^* = 5$ clusters such that $N_1 = 20$, $N_2 = 50$, $N_3 = 30$, $N_4 = 10$, $N_5 = 40$ across $p=10$ features, each with a proportion of explained variance of $E=0.6$ (top), $E=0.5$ (middle) or $E=0.4$ (bottom). Median and inter-quartile range of the calibrated number of clusters (G), Rand index, Adjusted Rand Index (ARI), Jaccard index, and computation time in seconds, as well as the percentage of significant clustering structure are reported. 
\pagebreak

\noindent
\renewcommand{\arraystretch}{0.8}
{\fontsize{10}{11}\selectfont
\begin{tabular}{cccccccc} \hline
n&Method&G&Rand index&ARI&Jaccard index&Percentage&Time (s)\\ \hline
&G*&5 [0]&0.924 [0.047]&0.792 [0.126]&0.727 [0.140]&&0 [0]\\
&Silhouette&4 [2]&0.887 [0.204]&0.710 [0.370]&0.648 [0.303]&&0 [0]\\
&GAP statistic&5 [0]&0.925 [0.039]&0.797 [0.109]&0.734 [0.123]&&6 [2]\\ \cline{2-8}
&G*&5 [0]&0.953 [0.034]&0.871 [0.091]&0.822 [0.113]&&11 [4]\\
300&Delta&3 [1]&0.740 [0.129]&0.472 [0.200]&0.473 [0.136]&&11 [4]\\
&PAC&20 [0]&0.768 [0.048]&0.318 [0.150]&0.309 [0.113]&&11 [4]\\
&PINS discrepancy&20 [15]&0.772 [0.118]&0.338 [0.416]&0.330 [0.345]&&11 [4]\\
&RCSI (PAC)&5 [1]&0.944 [0.076]&0.848 [0.179]&0.794 [0.196]&100\%&865 [507]\\
&RCSI (entropy)&5 [1]&0.945 [0.071]&0.852 [0.168]&0.798 [0.184]&100\%&825 [486]\\
&Consensus score&5 [1]&0.946 [0.075]&0.855 [0.174]&0.802 [0.189]&&11 [4]\\ \hline
&G*&5 [0]&0.927 [0.036]&0.800 [0.100]&0.737 [0.112]&&0 [0]\\
&Silhouette&4 [2]&0.876 [0.221]&0.697 [0.390]&0.638 [0.318]&&0 [0]\\
&GAP statistic&5 [0]&0.927 [0.033]&0.804 [0.089]&0.742 [0.103]&&16 [10]\\ \cline{2-8}
&G*&5 [0]&0.949 [0.064]&0.862 [0.155]&0.811 [0.174]&&33 [17]\\
600&Delta&3 [1]&0.730 [0.132]&0.457 [0.195]&0.465 [0.129]&&33 [17]\\
&PAC&5 [15]&0.777 [0.223]&0.404 [0.570]&0.387 [0.496]&&33 [17]\\
&PINS discrepancy&5 [15]&0.791 [0.229]&0.464 [0.573]&0.461 [0.499]&&33 [17]\\
&RCSI (PAC)&5 [1]&0.929 [0.098]&0.813 [0.221]&0.756 [0.228]&100\%&2636 [1912]\\
&RCSI (entropy)&5 [1]&0.931 [0.092]&0.819 [0.212]&0.762 [0.228]&100\%&2237 [1369]\\
&Consensus score&5 [2]&0.946 [0.117]&0.854 [0.252]&0.804 [0.257]&&33 [17]\\ \hline
\end{tabular}}
\vspace{10pt}

Supplementary Table 3: Clustering performances of (consensus) hierarchical clustering with different calibration strategies from $N=1,000$ simulated datasets corresponding to different numbers of items. We simulate $N=1,000$ datasets with $n=300$ (top) or $n=600$ (bottom) items split into $G^* = 5$ clusters such that $N_1 = 40$, $N_2 = 100$, $N_3 = 60$, $N_4 = 20$, $N_5 = 80$ (top) or $N_1 = 60$, $N_2 = 150$, $N_3 = 90$, $N_4 = 30$, $N_5 = 120$ (bottom) across $p=10$ features, each with a proportion of explained variance of $E=0.5$. Median and inter-quartile range of the calibrated number of clusters (G), Rand index, Adjusted Rand Index (ARI), Jaccard index, and computation time in seconds for hierarchical clustering with the simulated number of clusters ($G^*$), or calibrated by maximising the silhouette and GAP score, and for consensus hierarchical clustering with $G^*$ or calibrated using the $\Delta$, PAC, PINS discrepancy, RCSI and consensus scores are reported. The RCSI is computed using $N=25$ iterations. We also report the percentage of significant clustering structure for Monte Carlo approaches. For consensus clustering, the reported time includes both the computation of consensus matrices and the calibration procedure.

\noindent
\renewcommand{\arraystretch}{0.8}
{\fontsize{10}{11}\selectfont
\begin{tabular}{cccccccc} \hline
Model&Method&G&q&Rand index&ARI&Jaccard index&Time (s)\\ \hline
Hierarchical&G*&5 [0]&&0.851 [0.083]&0.597 [0.213]&0.534 [0.184]&0 [0]\\
&Silhouette&4 [3]&&0.810 [0.243]&0.515 [0.368]&0.473 [0.248]&0 [0]\\
&GAP statistic&5 [2]&&0.876 [0.062]&0.650 [0.180]&0.580 [0.178]&5 [1]\\ \hline
Unweighted&G*&5 [0]&&0.917 [0.085]&0.775 [0.207]&0.708 [0.226]&3 [1]\\
&Consensus score&5 [1]&&0.941 [0.071]&0.835 [0.189]&0.776 [0.214]&3 [1]\\ \hline
sparcl&$\lambda_1=10.00$&5 [0]&100 [0]&0.916 [0.087]&0.775 [0.212]&0.712 [0.231]&45 [17]\\
&$\lambda_2=7.83$&5 [0]&94 [4]&0.982 [0.089]&0.951 [0.220]&0.928 [0.275]&45 [17]\\
&$\lambda_3=6.12$&5 [0]&73 [8]&0.954 [0.114]&0.879 [0.273]&0.833 [0.319]&45 [17]\\
&$\lambda_4=4.79$&5 [0]&50 [7]&0.949 [0.117]&0.864 [0.277]&0.814 [0.312]&45 [17]\\
&$\lambda_5=3.75$&5 [0]&31 [6]&0.936 [0.111]&0.827 [0.269]&0.770 [0.297]&45 [17]\\
&$\lambda_6=2.93$&5 [0]&17 [3]&0.897 [0.115]&0.723 [0.284]&0.658 [0.289]&45 [17]\\
&$\lambda_7=2.30$&5 [0]&9 [2]&0.827 [0.111]&0.544 [0.279]&0.491 [0.236]&45 [17]\\
&$\lambda_8=1.80$&5 [0]&5 [1]&0.767 [0.090]&0.373 [0.230]&0.361 [0.157]&45 [17]\\
&$\lambda_9=1.41$&5 [0]&3 [0]&0.728 [0.073]&0.270 [0.178]&0.294 [0.110]&45 [17]\\
&$\lambda_10=1.10$&5 [0]&2 [0]&0.688 [0.061]&0.184 [0.113]&0.248 [0.059]&45 [17]\\
&Consensus score&3 [5]&82 [47]&0.660 [0.510]&0.372 [0.777]&0.413 [0.616]&45 [17]\\ \hline
COSA&$\lambda_1=10.00$&5 [0]&&0.951 [0.068]&0.865 [0.184]&0.813 [0.224]&1462 [115]\\
&$\lambda_2=5.99$&5 [0]&&0.961 [0.060]&0.893 [0.169]&0.849 [0.213]&1462 [115]\\
&$\lambda_3=3.59$&5 [0]&&0.969 [0.047]&0.915 [0.130]&0.879 [0.171]&1462 [115]\\
&$\lambda_4=2.15$&5 [0]&&0.976 [0.039]&0.935 [0.108]&0.905 [0.146]&1462 [115]\\
&$\lambda_5=1.29$&5 [0]&&0.974 [0.040]&0.930 [0.108]&0.898 [0.147]&1462 [115]\\
&$\lambda_6=0.77$&5 [0]&&0.963 [0.043]&0.897 [0.118]&0.855 [0.151]&1462 [115]\\
&$\lambda_7=0.46$&5 [0]&&0.931 [0.061]&0.809 [0.170]&0.747 [0.189]&1462 [115]\\
&$\lambda_8=0.28$&5 [0]&&0.868 [0.081]&0.635 [0.206]&0.566 [0.188]&1462 [115]\\
&$\lambda_9=0.17$&5 [0]&&0.780 [0.096]&0.403 [0.224]&0.377 [0.151]&1462 [115]\\
&$\lambda_10=0.10$&5 [0]&&0.681 [0.087]&0.182 [0.154]&0.249 [0.076]&1462 [115]\\
&Consensus score&5 [1]&&0.977 [0.032]&0.936 [0.085]&0.908 [0.114]&1462 [115]\\ \hline
\end{tabular}}
\vspace{10pt}

Supplementary Table 4: Comparison of clustering performances of (consensus) clustering using the Euclidean distance (unweighted), sparcl or COSA. Performances are evaluated on $N=1,000$ datasets with $n=150$ items split into $G^* = 5$ clusters such that $N_1 = 20$, $N_2 = 50$, $N_3 = 30$, $N_4 = 10$, $N_5 = 40$ across $p=100$ features, of which $q^*=20$ have a nonzero proportion of explained variance ($E=0.6$). Median and inter-quartile range of the calibrated number of clusters (G), Rand index, Adjusted Rand Index (ARI), Jaccard index, and computation time in seconds for hierarchical clustering with the simulated number of clusters ($G^*$), or calibrated by maximising the silhouette and GAP score, and consensus hierarchical, sparcl and COSA clustering with $G^*$ for different values of $\lambda$ (sparcl or COSA only) and calibrated by maximising the consensus score are reported.
\pagebreak

\end{document}